\documentclass[usenatbib,useAMS english]{mnras}
\usepackage[fleqn]{amsmath}
\usepackage{array,multirow,graphicx}
\usepackage{txfonts}
\usepackage{lmodern}
\bibpunct{(}{)}{;}{a}{}{,}
\usepackage{floatrow}
\usepackage{array}
\usepackage{relsize}
\usepackage{bm}
\usepackage{listings}
\usepackage{url}
\usepackage[font=small,labelfont=bf]{caption}

\graphicspath{{./figs/}}

\newcommand{\HI}{H\,{\sc i} } 
\newcommand{\HInospace}{H\,{\sc i}}
\newcommand{\Htwo}{H$_{2}$ }

\newfloatcommand{capbtabbox}{table}[][\FBwidth]

\def\app#1#2{%
\mathrel{%
\setbox0=\hbox{$#1\sim$}%
\setbox2=\hbox{%
\rlap{\hbox{$#1\propto$}}%
\lower1.1\ht0\box0%
}%
\raise0.25\ht2\box2%
}%
}

\title[Physical drivers of galaxies' cold-gas content with {\sc Dark Sage}]{Physical drivers of galaxies' cold-gas content: exploring environmental and evolutionary effects with {\sc Dark Sage}}
\author[A.~R.~H.~Stevens \& T. Brown]{Adam R.~H.~Stevens$^{1,2}$\thanks{E-mail: adam.stevens@uwa.edu.au} and Toby Brown$^{1,2}$\\
$^1$Centre for Astrophysics and Supercomputing, Swinburne University of Technology, Hawthorn, VIC 3122, Australia\\
$^2$International Centre for Radio Astronomy Research, The University of Western Australia, Crawley, WA 6009, Australia}

\begin{document}

\pagerange{\pageref{firstpage}--\pageref{lastpage}} \pubyear{2017}

\maketitle

\label{firstpage}

\begin{abstract}
We combine the latest spectrally stacked data of 21-cm emission from the ALFALFA survey with an updated version of the {\sc Dark Sage} semi-analytic model to investigate the relative contributions of secular and environmental astrophysical processes on shaping the \HI fractions and quiescence of galaxies in the local Universe.  We calibrate the model to match the observed mean \HI fraction of all galaxies as a function of stellar mass.  Without consideration of stellar feedback, disc instabilities, and active galactic nuclei, we show how the slope and normalisation of this relation would change significantly.  We find {\sc Dark Sage} can reproduce the \emph{relative} impact that halo mass is observed to have on satellites' \HI fractions and quiescent fraction. However, the model satellites are \emph{systematically} gas-poor.  We discuss how this could be affected by satellite--central cross-contamination from the group-finding algorithm applied to the observed galaxies, but that it is not the full story.  From our results, we suggest the anti-correlation between satellites' \HI fractions and host halo mass, seen at fixed stellar mass and fixed specific star formation rate, can be attributed almost entirely to ram-pressure stripping of cold gas.  Meanwhile, stripping of hot gas from around the satellites drives the correlation of quiescent fraction with halo mass at fixed stellar mass.  Further detail in the modelling of galaxy discs' centres is required to solidify this result, however.  We contextualise our results with those from other semi-analytic models and hydrodynamic simulations.
\end{abstract}


\section{Introduction}
\label{sec:intro}
It is widely accepted that the seeds of galaxies are born out of the gravitational fragmentation of gas within dark-matter-dominated overdensities in the Universe \citep[\'{a} la][]{white78}, more commonly referred to as `haloes'.  Our modern theoretical consensus is that galaxies then continue to accrete gas cosmologically, either directly through cold streams or via their hot circumgalactic medium, where ionised gas must cool to a predominantly neutral phase before settling into the galactic disc \citep[see][]{rees77,white91,birnboim03,keres05,dekel06,benson11,stevens17}.  In addition, the hierarchical nature of galaxy growth means galaxies can acquire large volumes of gas at once through mergers.  As the principal component of gas in galaxies, and because cold gas is the raw material for forming new stars, neutral atomic hydrogen (\HInospace) is one of the most important ingredients in the formation and evolution of galaxies.  Therefore, understanding the ensemble of internal and external astrophysical mechanisms that regulate \HI content and star formation activity is critical if we are to realise a coherent picture of galaxy evolution.

Internally, beyond the direct consumption of gas in the formation of stars and the accretion of black holes, the evolution of galaxies' gas reservoirs is consequently dictated by feedback from stellar evolution and active galactic nuclei.  This feedback manifests in the form of energetic winds that eject and/or heat gas reserves, as evidenced and detailed by a variety of observations \citep[e.g.][]{sharp10,feruglio10,anderson15,cicone15,nielsen16} and simulations \citep*[e.g.][]{springel03,brook11,costa15,taylor15,bower17}. Instabilities and turbulence within the interstellar medium also help regulate the distribution of gas and its ability to form stars \citep*[see, e.g.,][]{federrath15,stevens16}.  While these secular processes all impact galaxies' gas fractions, it has become increasingly clear that external effects, driven by environment, are of comparable importance.

The role of environment in the suppression of \HI was first demonstrated by observations showing cluster galaxies to be gas-poor compared to the field \citep{giovanelli85,solanes01}. This was exemplified in exquisite detail by the VLA\footnote{Very Large Array} Imaging of Virgo in Atomic gas survey \citep[VIVA;][]{chung09}, which demonstrated the importance of the highest density environments in shutting down star formation via strong gas depletion mechanisms. Other observational \citep[e.g.][]{cortese11,catinella13} and theoretical \citep[e.g.][]{mccarthy08,rafie15,marasco16} efforts support this picture, with both camps generally describing external processes that are distinguishable based upon the time-scales over which they act: (i) those that act swiftly and directly upon the cold gas to remove it from the galaxy via an interaction of the interstellar and intracluster (or intragroup) media \citep[$\sim$10s of Myr, i.e. ram-pressure stripping;][]{gunn72}, or (ii) those that regulate the rate at which gas is able to accrete onto the galaxy from its dark-matter halo over more lengthy time-scales \citep*[i.e.~strangulation, where galaxies consume their available gas for star formation in $\gtrsim$1 Gyr;][]{larson80}. Using this distinction, \citet{brown17} provide strong evidence that the gas loss in massive haloes ($M_{\rm vir} > 10^{13}\,{\rm M}_{\odot}$) is considerably faster than the subsequent quenching of star formation.

To zeroth order, environment can be thought of as a dichotomy between `central' and `satellite' galaxies.  These terms are most relevant for numerical simulations, where a central galaxy belongs to the most massive subhalo of a halo \citep[see, e.g.,][]{springel01,onions12}.  All remaining subhaloes host satellite galaxies.  Under this definition, a central can be both isolated or have any number of satellites associated with it.  By definition, only satellite galaxies are prone to gas-stripping processes.  For observational data to be interpreted in this framework, group-finding algorithms are typically employed to arrange galaxies into haloes.  These use a known halo mass function from simulations with abundance matching to assign halo masses, where the brightest or most massive galaxy is labelled the central and the remainder are tagged as satellites \citep[see, e.g.,][]{huchra82,tucker00,yang05,campbell15}.  The strength of environmental stripping processes depends primarily on the parent halo mass, within which the satellite resides, as evidenced by their \HI fractions \citep{brown17}.  

In recent years, single-dish surveys have begun to provide \HI measurements for tens of thousands of galaxies. For example, the \HI Parkes All Sky Survey \citep[HIPASS;][]{meyer04} scanned $\sim$75 per cent of the sky (30,000 deg$^2$) and detected 21-cm emission in $\sim$5000 nearby galaxies. Its successor, the Arecibo Legacy Fast ALFA\footnote{Arecibo L-band Feed Array} survey \citep[ALFALFA;][]{giovanelli05} went even further, and the latest data release ($\alpha$.70) provides a census of \HI gas for $\sim$25\,000 galaxies over $\sim$7000 deg$^2$ of sky \citep[for a presentation of the earlier $\alpha$.40 release, see][]{haynes11}. In addition, the stacking of undetected \HI sources based on optical position and redshift has become be a major tool for pushing 21-cm surveys beyond their nominal sensitivity limit, providing representative, statistical studies of \HI as a function of galaxy properties and environment in the local Universe \citep{fabello11a,fabello11b,fabello12,brown15,brown17}.

In their paper, \citet{brown17} compare their average gas content scaling relations with predictions from the cosmological hydrodynamic simulations of \citet{dave13} and the {\sc galform} semi-analytic model \citep{gp14}. They find that both sets of predictions are too gas-poor at fixed stellar mass and specific star formation rate (sSFR), although there is qualitative reproduction of ram-pressure stripping in the group and cluster regime. \citet{marasco16} also investigate the role of environment in dictating the \HI content of galaxies within the EAGLE\footnote{Evolution and Assembly of GaLaxies and their Environments} suite of cosmological hydrodynamic simulations \citep{schaye15}. They use the simulations to successfully reproduce the findings of \citet{fabello12} and \citet{catinella13}, demonstrating that EAGLE galaxies in more massive haloes have lower gas content at fixed stellar mass. However, rather than a continuous trend of \HI depletion as a function of environment \citep[e.g.][]{stark16,brown17}, their work finds the role of environment to be more binary, determining whether or not galaxies have any \HI at all.  Higher-resolution runs of these simulations have found the \HI fractions to increase \citep{bahe16,marasco16,crain17}.

Despite significant progress, large uncertainties remain concerning where and how precisely, in terms of epoch and environment, secular and environmental processes influence galaxy gas reservoirs. We therefore use the {\sc Dark Sage} semi-analytic model \citep{stevens16} to investigate the contributions of various evolutionary and environmental processes in determining the gas fractions and star formation activity of both central and satellite galaxies, relating this to their stellar masses and specific star formation rates.  Because semi-analytic models describe the phenomenology of hydrodynamical effects without directly modelling hydrodynamics at all, all gas-stripping processes are manually prescribed.  The advantage this gives is that we can learn about the relative roles each effect has on the gas content of galaxies by turning the effects on and off.  {\sc Dark Sage} is especially well equipped, as the one-dimensional disc structure of galaxies is self-consistently evolved in the model, naturally leading to a spatial dependence of environmental effects within galaxies.

This paper is structured as follows.  We describe the main aspects of the {\sc Dark Sage} semi-analytic model in Section \ref{sec:sam}, including new features we have introduced for this work.  In Section \ref{sec:obs}, we describe the observational data with which our results are closely compared.  We examine the \HI fractions and quiescence (lack of star formation) of galaxies within the central--satellite dichotomy in Section \ref{sec:satcen} by comparing a set of {\sc Dark Sage} model variants against our observed data.  The role of environment in shaping the \HI fractions and quiescence of satellite galaxies is then studied in greater detail in Section \ref{sec:satenv}.  We discuss the context of our results with recent literature and offer concluding remarks in Section \ref{sec:discussion}.

\section{The semi-analytic model}
\label{sec:sam}

The {\sc Dark Sage} semi-analytic model \citep{stevens16} is a heavily modified version of the models developed by \citet{croton06,sage}.  Whereas most semi-analytic models evolve the integrated properties of singular baryonic reservoirs, {\sc Dark Sage} evolves the one-dimensional structure of galactic discs in annuli of fixed specific angular momentum \citep[similar to][]{stringer07,dutton09}, where $j \! = \! r \, v_{\rm circ}$.  Evolutionary and environmental processes thus affect each galaxy on local scales.  We refer the reader to section 3 of \citet{stevens16} for full details, but briefly outline the key features of the model here.

\begin{itemize}
\item \emph{Cooling:} Hot gas cools onto galaxies following the method introduced by \citet{white91}, using the cooling tables of \citet{sutherland93}.  The cooling gas is assumed to have an exponential surface density profile as a function of $j$ and to immediately localise itself with material in the disc of the same $j$.

\item \emph{Star formation, evolution, and feedback:} Passive star formation goes as $\Sigma_{\rm SFR} \! \propto \! \Sigma_{\rm H_2}$, where 43\% of star-forming gas is approximated as instantly recycled \citep[cf.][]{cole00}.  Every star formation episode results in supernova feedback, which heats gas out of the disc.  The mass of gas heated in an annulus is directly proportional to the mass of stars formed and inversely proportional to the local gas surface density.  The latter approximates supernova energy as being more readily dissipated in denser gaseous regions \citep[this model follows][]{fu10,fu13}.  Excess energy from supernovae can eject gas out of the halo.

\item \emph{Instabilities:} A two-component Toomre $Q$ value \citep{toomre64,romeo11} for each annulus is routinely calculated.  If $Q<1$, a starburst in that annulus can occur, and any remaining unstable mass is transferred to the two adjacent annuli in proportion such that angular momentum is conserved.  Sinking unstable stars in the innermost annulus are transferred to a fully dispersion-dominated bulge component, while gas can be accreted onto the central black hole.  

\item \emph{Mergers:} For minor mergers, the specific angular momentum of the merging satellite is measured \emph{relative to the central galaxy} to determine where in the disc the satellite's cold gas ends up, while its stars are transferred to the central's bulge.  For major mergers, the gas disc vectors are summed to define the new plane onto which both gas discs are projected, while both stellar discs are destroyed and form a bulge.  Starbursts occur in the annuli where gas originated from both systems, based on a variation of \citet*{somerville01}.  Black holes are directly fed gas during mergers phenomenologically \citep[following][]{kauffmann00}.

\item \emph{Active galactic nuclei:} Radio mode accretion of hot gas onto a central black hole and subsequent feedback occurs at each time-step, whereby hot gas is prevented from cooling \citep{sage}.  Quasar mode feedback is triggered by mergers and instabilities.  The model steps out annulus by annulus, determines whether the quasar energy is sufficient to strip the cumulative gas up to that annulus, and if so, transfers all the gas in that annulus to the hot reservoir.  The remaining energy can eject gas out of the halo entirely.
\end{itemize}

By default, whenever gas is reheated by feedback, it is transferred to the hot and/or ejected reservoirs associated with the \emph{central} galaxy within the halo, even if the gas originated from a satellite (but see Section \ref{ssec:hot}).  While the choice of whether reheated gas remains in a satellite's hot reservoir or not can have some effect on the gas properties of model galaxies \citep[e.g., as raised by][]{font08,lagos14}, this effect is entirely negligible for the results we present for {\sc Dark Sage}.

{\sc Dark Sage} has been run on the standard merger trees of the Millennium simulation \citep{millennium}.  The simulation included $2160^3$ particles of mass $1.18 \! \times \! 10^9\, {\rm M}_{\odot}$ in a periodic box with a comoving length of 685\,Mpc.  This assumes $h\!=\!0.73$, $\Omega_M\!=\!0.25$, $\Omega_\Lambda\!=\!0.75$, $\Omega_b\!=\!0.045$, and $\sigma_8\!=\!0.9$ \citep{wmap1}.  The merger trees carry 64 snapshots.  As in {\sc sage} \citep{sage}, galaxies in {\sc Dark Sage} are evolved on 10 sub-time-steps between each snapshot.

Galaxy catalogues built from {\sc Dark Sage} are available through the Theoretical Astrophysical Observatory \citep{tao}.\footnote{\url{https://tao.asvo.org.au}}  Here, photometric properties and light-cones for {\sc Dark Sage} galaxies can be constructed.  The codebase itself has now also been made publicly available.\footnote{\url{https://github.com/arhstevens/DarkSage}}

\subsection{Updates to the model}
For the purposes of this work, we have updated some of the default prescriptions in {\sc Dark Sage} and added a few optional variants to the code as well.  These were added to improve the average gas content of the entire galaxy population at low stellar masses (see Section \ref{ssec:recal}) and to increase our ability to investigate the impact of environment on gas content (see below).

\subsubsection{Atomic and molecular hydrogen}
\label{ssec:hih2}
The amount of gas in the form of \HI and H$_2$ in each annulus of each {\sc Dark Sage} galaxy is regularly recalculated.  In \citet{stevens16}, this was based on the mid-plane pressure of the disc \citep{blitz04}, and closely followed the prescription of \citet{fu10}.  While we include that prescription in this work, we have now opted for a metallicity-based prescription as our default, where
\begin{subequations}
\label{eq:RH2}
\begin{equation}
R_{{\rm H}_2}(r) \equiv \frac{\Sigma_{{\rm H}_2}(r)}{\Sigma_{\rm H\,\textsc{i}}(r)} = \left[ \left( 1 - \frac{3\,s(r)}{4+s(r)}\right)^{-1} - 1 \right]^{-1}~,
\end{equation}
\begin{equation}
s(r) \equiv \frac{\ln\left(1+0.6\chi(r) +0.01\chi^2(r) \right)}{0.6 \tau(r)}~,
\end{equation}
\begin{equation}
\chi(r) \equiv \frac{3.1}{4.1} \left[ 1 + 3.1 \left(\frac{Z(r)}{{\rm Z}_{\odot}}\right)^{0.365} \right]~,
\end{equation}
\begin{equation}
\tau(r) \equiv 0.66\, \frac{Z(r)}{{\rm Z}_{\odot}} \frac{c_f(r)\, \Sigma_{\rm gas}(r)}{{\rm M}_{\odot}\, {\rm pc}^{-2}}~,
\end{equation}
\begin{equation}
c_f(r) \equiv \phi \left(\frac{Z(r)}{{\rm Z}_{\odot}}\right)^{-\gamma}~,
\end{equation}
\end{subequations}
where $\Sigma(r)$ is the local surface density of the subscripted quantity [where $\Sigma_{\rm gas}(r)$ is for \emph{all} gas in the disc], $Z(r)$ is the local surface density ratio of gaseous metals to all gas, and $c_f(r)$ is the local clumping factor.  This is based on the work of \citet{mckee10} and follows the implementation of \citet{fu13}.  These equations originate from modelling spherical gas clouds subject to a photoionizing ultra-violet background, where metallicity represents the dust content of those clouds, which shield the cloud from the ionizing photons, allowing a greater fraction of the gas to be in molecular form \citep*[for further details, see][]{krumholz08,krumholz09}.  Here, we assume a solar metallicity of ${\rm Z}_{\odot} \! = \! 0.02$ and treat $\phi$ and $\gamma$ as free parameters, for which we find values of 3.0 and 0.3, respectively, after recalibrating the model (Section \ref{ssec:recal}).

Note that $R_{\rm H_2}(r)$ can only be calculated by equation (\ref{eq:RH2}) provided $s(r) \! < \! 2$.  This requires gas to have non-zero metallicity.  In {\sc Dark Sage}, gas begins with zero metallicity.  If passive star formation from H$_2$ were the only means of forming stars in the model, {then we would have a paradoxical chicken--egg scenario with neither chickens nor eggs,} and hence star formation would never ignite with this method.  This is resolved by instabilities and mergers, which both induce starbursts independent of any H$_2$ (for full details, see \citealt{stevens16}).  In other words, once a galaxy has experienced an instability or merger, which tends to happen quickly, it can form stars regularly from H$_2$ via equation (\ref{eq:RH2}).

In this paper, we show results from {\sc Dark Sage} using both the metallicity- and pressure-based prescriptions for determining the ratio of H$_2$ (and \HInospace) in gas disc annuli.  We denote these as $f_{\rm H_2}(Z)$ and $f_{\rm H_2}(P)$, respectively, where $f_{\rm H_2}$ is formally defined as the mass ratio of H$_2$ to \emph{all} gas within an annulus, as per equation 12 of \citet{stevens16}, and hence has a one-to-one mapping with $R_{\rm H_2}$.

\begin{figure}
\includegraphics[width=\linewidth]{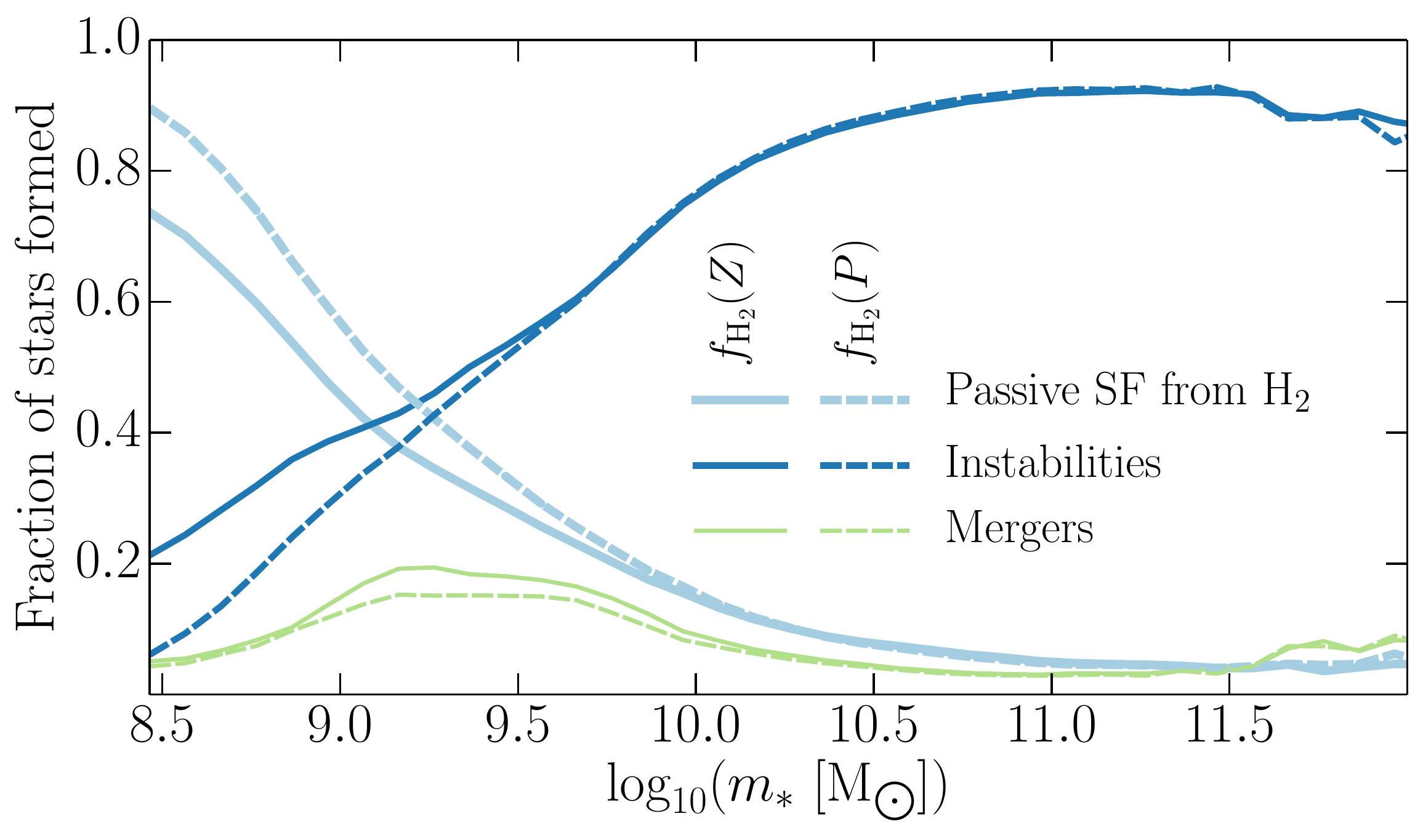}
\caption{Fraction of stars formed through the three possible channels in {\sc Dark Sage} as a function of galaxy stellar mass at $z\!=\!0$.  This accounts for both in-situ and ex-situ star formation.  Galaxies are grouped in 0.1-dex-wide bins.  Solid curves use the metallicity-based prescription for determining the ratio of atomic and molecular hydrogen in gas disc annuli.  Dashed curves use the mid-plane pressure prescription.  Both include the full model's physics otherwise.}
\label{fig:sfchannels}
\end{figure}

By changing the prescription for $f_{\rm H_2}$, the relative contributions of each star formation channel to the final stellar populations at $z\!=\!0$ are altered for galaxies of mass $m_* \! \lesssim \! 10^{10}\,{\rm M}_{\odot}$.  This is demonstrated in Fig.~\ref{fig:sfchannels}.  Compared to $f_{\rm H_2}(P)$, switching to $f_{\rm H_2}(Z)$ reduces the contribution of the passive H$_2$ channel of star formation, and increases the contributions from both instability- and merger-driven starbursts.  Regardless of the $f_{\rm H_2}$ prescription though, the stars in galaxies of mass $m_* \! \lesssim \! 10^9\,{\rm M}_{\odot}$ are predominantly born out of the H$_2$-dependent passive channel of star formation, whereas the stars in galaxies of mass $m_* \! \gtrsim \! 10^{9.5}\,{\rm M}_{\odot}$ are mainly born in instability-driven starbursts.  This has important consequences for some of the runs of the model we present in this paper, which we come back to in Section \ref{ssec:evolution}.

\subsubsection{Ram-pressure stripping of cold gas}
\label{ssec:rps}
As was the case in \citet{stevens16}, the cold gas in each annulus of each satellite galaxy is subject to ram-pressure stripping from the hot gas associated with the central (most massive) subhalo, based on the landmark work of \citet{gunn72}:
\begin{equation}
\rho_{\rm hot, cen}(R_{\rm sat})\, v_{\rm sat}^2 \geq 2 \pi G\, \Sigma_{\rm gas}(r) \left[\Sigma_{\rm gas}(r) + \Sigma_*(r) \right]~,
\label{eq:rps}
\end{equation}
where $\rho_{\rm hot, cen}(R_{\rm sat})$ is the local density of the intracluster medium (or the circumgalactic medium of the central) at the position of the satellite and $v_{\rm sat}$ is the relative speed of the satellite to the central.  If this inequality is met for an annulus, all gas in that annulus is transferred to the hot reservoir of the central.  

However, equation (\ref{eq:rps}) does not provide a criterion for \emph{when} a disc will feel ram pressure; it simply tells us \emph{if} it is vulnerable to ram pressure how much gas will be stripped.  In order for a disc to be vulnerable, the satellite's hot gas on its leading side must first be stripped away.  While the stripping of a satellite's hot gas would be asymmetric in reality, {\sc Dark Sage} has built-in symmetry assumptions about matter in (sub)haloes (like most semi-analytic models).  Bearing all of this in mind, we have added a condition to {\sc Dark Sage} that equation (\ref{eq:rps}) is only applied to satellites if the total baryonic mass (cold gas + stars) of the galaxy exceeds the hot-gas mass of the subhalo.  While this phenomenological condition misses a lot of the complex physics surrounding ram-pressure stripping, it provides a simple means of ensuring that some of a satellite's protective hot gas must be lost before cold gas in the disc is susceptible to stripping.  This is more important for massive satellites \citep[see, e.g., fig.~2 of][]{sage}.

\subsubsection{Stripping of hot gas in subhaloes}
\label{ssec:hot}
As an alternative to hot gas being stripped in proportion to dark matter \citep[see][]{sage}, which we maintain as the default for {\sc Dark Sage}, we have now included an option for a ram-pressure-stripping-like prescription of the hot gas in subhaloes \citep[\`{a} la][]{font08,mccarthy08}.  Specifically, we assume equation 6 of \citet{mccarthy08} describes the gravitational restoring force per unit area on the hot gas of the satellite, which is assumed to be distributed in a singular isothermal sphere.  We then find the radius $R$ from the satellite's centre where
\begin{equation}
\rho_{\rm hot, cen}(R_{\rm sat})\, v_{\rm sat}^2 = \frac{G\,M_{\rm sat}(<\!R)\,m_{\rm hot,sat}}{8\,R_{\rm vir,sat}\,R^3}~,
\label{eq:rps_hot}
\end{equation}
which equates ram pressure with the gravitational restoring force density (where subscript `sat' implies the subhalo the satellite is associated with, and $M_{\rm sat}$ accounts for all matter).  Any hot gas in the subhalo external to this radius is assumed to be stripped over the course of the simulation snapshot interval, meaning one tenth of that mass is stripped in a sub-time-step on which the galaxies are evolved.  For the sake of simplicity, after each stripping calculation, the remaining hot gas is assumed to redistribute itself into a singular isothermal sphere again, such that equation (\ref{eq:rps_hot}) is always applicable.  If left unchecked, this artificial redistribution of hot gas to larger radii would lead to continually stronger stripping on each consecutive sub-time-step.  To counteract this, the stripped hot-gas mass at a given sub-time-step is not allowed to exceed that of the first sub-time-step within the same snapshot interval.  Once the next snapshot is reached, the virial radius of the subhalo will have reduced, and hence so will the extent to which the hot gas is assumed to be distributed.

We also include the option of not stripping hot gas at all.  With this option, we also return reheated gas from feedback to the hot component of the subhalo, rather than that of the main halo, which would otherwise be standard.

\subsection{Recalibration}
\label{ssec:recal}
\begin{figure}
\centering
\includegraphics[width=\linewidth]{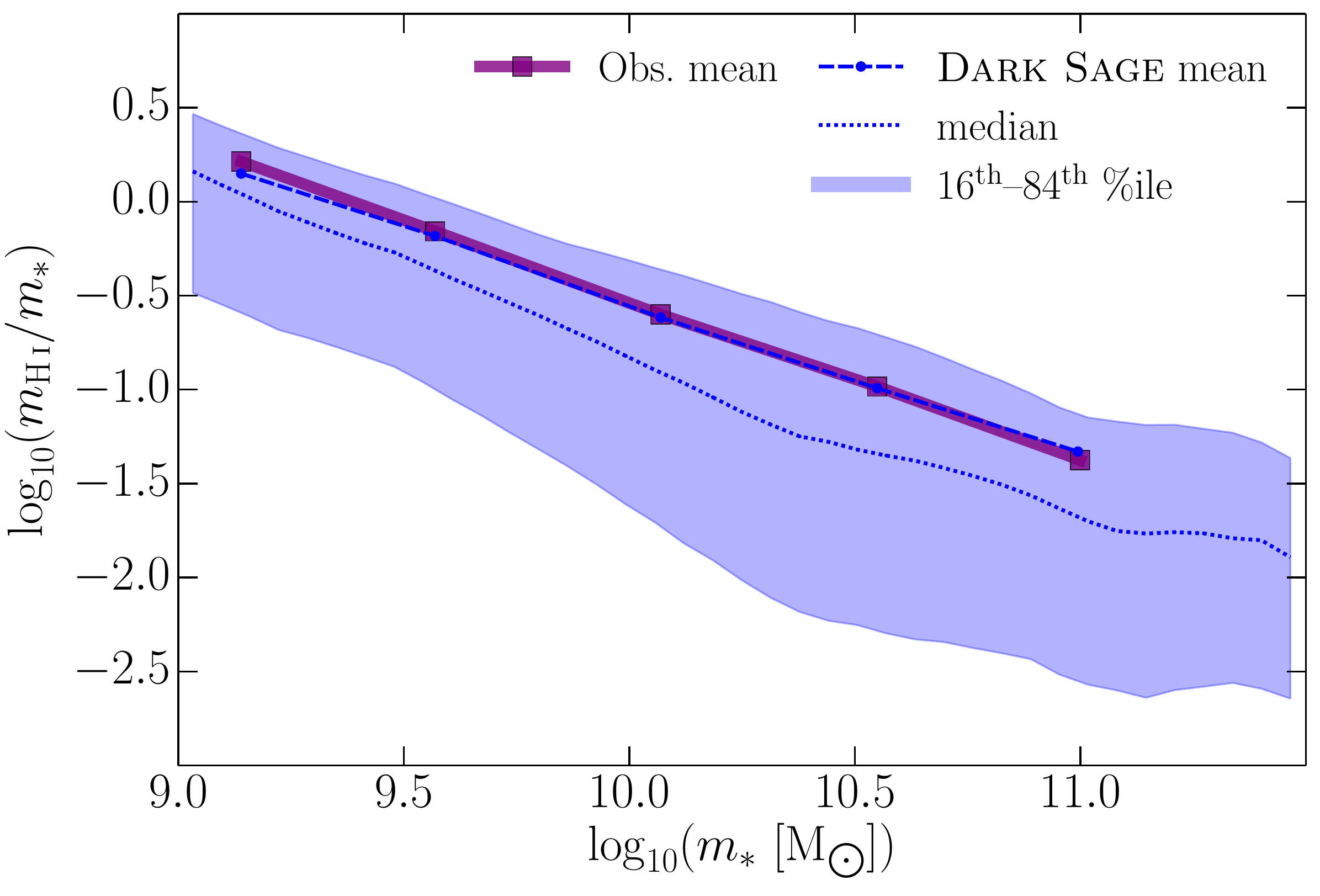}
\caption{\HI fraction of {\sc Dark Sage} galaxies at $z\!=\!0$ as constrained against the observations of \citet{brown15}.  The points connected by solid lines give the logarithm of the mean along each axis for five bins, for each of the observed and model galaxies.  The dashed curve and shaded region give the median \HI fraction and 16$^{\rm th}$--84$^{\rm th}$ percentile range of the model galaxies, respectively.}
\label{fig:hifrac}
\end{figure}

Before investigating how environmental and evolutionary processes can affect the gas fractions of galaxies, we first need to ensure the average gas fraction of {\sc Dark Sage} galaxies is representative of the real Universe.  This was always a constraint used in calibrating the model \citep[using the data from][]{brown15}, but there was still an overall deficit in the gas content of galaxies at low stellar mass \citep[see appendix A of][]{stevens16}.  This has now been alleviated through a combination of the updates to the code above and opting for the \HI fraction of galaxies to become the constraint of greatest emphasis.  We further exclude any subhaloes that were never composed of at least 100 particles in their history, ensuring that we use well-resolved galaxies in calibrating the model.  This is now in much finer agreement with \citet{brown15}, as presented in Fig.~\ref{fig:hifrac}.

The other $z\!=\!0$ constraints of the model have been maintained as well.  These include the stellar, H\,{\sc i}, and H$_2$ mass functions, the black hole--bulge mass relation, the stellar mass--gas metallicity relation, and the Baryonic Tully--Fisher relation.  See section 3 and appendix A of \citet[][and references therein]{stevens16} for further details.

We have only modified one parameter from the original version of {\sc Dark Sage}; we have reduced the passive star formation efficiency from H$_2$ by a factor of 3 to $\epsilon_{\rm SF} \! = \! 1.3 \! \times \! 10^{-4}\,{\rm Myr}^{-1}$ (cf.~table 1 of \citealt{stevens16}).  This means a greater fraction of stars are formed through disc instabilities.  Specifically, with the old parameter value, the dominant star formation channel transitioned from passive H$_2$ to instabilities at a factor of $\sim$10 higher in stellar mass versus what is seen now in Fig.~\ref{fig:sfchannels}.  This has a minimal effect on the stellar mass function, but it is important for when stars form in a galaxy, and for their total \HI and H$_2$ content.  This parameter change has been implemented for \emph{all} runs of the model presented in this paper.

\subsection{Model galaxy sample}
\label{ssec:sample}
Throughout this paper, we study galaxies from {\sc Dark Sage} at $z\!=\!0$, and only include those with $10^9 \! < \! m_*/{\rm M}_{\odot} \! \leq \! 10^{11.5}$, matching our sample of observed galaxies, described in Section \ref{sec:obs}.  In principle, we could have used the $z\!=\!0.041$ snapshot from Millennium, which would have more accurately matched the redshift range for the observations.  However, the evolution of galaxies is sufficiently low at $z \! \lesssim \! 0.05$ such that our results would have no noticeable changes.

We note that while the \HI and stellar mass measurements we present for {\sc Dark Sage} galaxies are instantaneous quantities for the $z\!=\!0$ snapshot, star formation rates are average quantities over the period since the previous snapshot.  The time-scale for the star formation rates is hence $\sim$260\,Myr.  While this is an order-of-magnitude difference in time-scale from the H$\alpha$ measurements for the observed galaxies (cf.~Section \ref{sec:obs} of this paper and section 6 of \citealt{brown17}), it is not a cause of significant concern.  \citet{benson12} has shown that improving the temporal resolution of a simulation above Millennium's (64 snapshots) will only result in a $\lesssim$5 per cent change in the universal average SFR from the {\sc galacticus} semi-analytic model \citep{galacticus} at any time.  The same is true for the stellar mass function at $z \! \lesssim \! 4$ \citep[see fig.~11 of][]{benson12}.  \citet[][see appendix B]{thesis} showed similar results for {\sc sage} \citep{sage} using the GiggleZ simulation suite \citep{gigglez}.  Using the same simulations, we have confirmed that any sSFR-related results for {\sc Dark Sage} are negligibly affected by temporal resolution, whether we increase it by a factor of 4 or decrease it by a factor of 2 (not shown here).
\section{Observational data}
\label{sec:obs}

This section describes the observational data used in this paper. Galaxies are selected according to their stellar mass ($m_* \! \geq \! 10^9 \, {\rm M}_{\odot}$) and redshift ($0.02 \! \leq \! z \! \leq \! 0.05$) from the overlap in volume between the Sloan Digital Sky Survey Data Release 7 \citep[SDSS DR7;][]{abazajian09} and the ALFALFA survey \citep{giovanelli05}. This yields a representative parent sample of 30,695 galaxies that is defined in Section 2 of \citet{brown15}.  We refer the reader to that paper for a more detailed description.

\HI 21-cm spectra are extracted from ALFALFA data cubes using the SDSS DR7 coordinates and redshift. This provides spectra for every galaxy in the sample, regardless of whether it is a formal \HI detection or not.

The MPA-JHU SDSS DR7 catalogue\footnote{\url{http://www.mpa-garching.mpg.de/SDSS/DR7/} and improved stellar masses from \url{http://home.strw.leidenuniv.nl/~jarle/SDSS/}}
provides stellar mass and star formation rate (SFR) estimates for the entire sample. Masses are constrained following the methodology of \citet{kauffmann03} using stellar absorption lines and fits to the optical photometry. The catalogue follows \citet{brinchmann04} in deriving global SFRs by using fits to emission H$\alpha$ line fluxes and, where no or low-signal-to-noise lines are detected, the 4000-\AA~break strength.  We note these probe completely different time-scales, but, as raised in Section \ref{ssec:sample}, both can be safely compared with our model results.

Halo masses are assigned using the SDSS DR7 group catalogue,\footnote{\url{http://gax.shao.ac.cn/data/Group.html}} detailed in \citet{yang07}. We refer the reader to their paper for a full description. Briefly, the authors use the friends-of-friends, iterative group-finding algorithm of \citet{yang05} to identify galaxy groups with SDSS DR7. Groups are then assigned a halo mass via the abundance matching technique, whereby individual group luminosity or stellar mass is ranked and then matched to the halo mass function of \citet{warren06}. For this work, we use the halo masses estimated using the stellar mass rank order.

For a full breakdown of the optical and \HI data, see sections 2.1 and 2.2 of \citet{brown15}. The sample's stellar mass, SFR, and halo mass estimates are described in more detail in section 2 of \citet{brown17}.  The galaxy and halo properties presented in those papers assumed a \citet{kroupa01} initial mass function (IMF) and a dimensionless Hubble parameter $h\!=\!0.7$.  In order to be consistent with the model galaxies we compare against (see Section \ref{sec:sam}), all quantities have been converted to assume $h\!=\!0.73$ and a \citet{chabrier03} IMF. We have assumed $0.66\, m_{\rm *, Chabrier}\!=\!0.61\, m_{\rm *, Kroupa}$ and $0.67\, {\rm SFR}_{\rm Chabrier}\!=\!0.63\, {\rm SFR}_{\rm Kroupa}$ \citep[as in][]{madau14}.

\subsection{\HI spectral stacking}
\label{ssec:stacking}

As outlined in the Section \ref{sec:intro}, \HI spectral stacking improves the statistical size of studies way beyond what is currently possible using only detections, probing the gas content of galaxies across the entire gas-rich to -poor regime. The stacking technique in this paper is based upon the technique developed by \citet{fabello11a} and described fully in section 3 of \citet{brown15}. To briefly summarise, we shift the \HI spectra of a given ensemble of galaxies to their rest frame velocity and co-add them regardless of their detection status, weighting each by the inverse of their corresponding stellar mass. Doing so yields a weighted average stacked profile and, therefore, the average \HI fraction, $\langle m_{\mathrm{H}\,\LARGE\textsc{i}} / m_* \rangle$, for the galaxies in each stack.

Errors on the stacking results are estimated using a statistical {\it delete-a-group jackknife} routine whereby a random 20 per cent of spectra in a given stack are iteratively discarded without replacement, each time re-estimating the average gas fraction. This provides five separate estimates of gas fraction with each spectra discarded once. Following equation 3 in \citet{brown15}, the error is then calculated as the standard deviation of these five average gas fraction measurements multiplied by a $\sqrt{N-1}$ factor, where $N\!=\!5$.
\section{Satellites versus centrals}
\label{sec:satcen}

\begin{figure*}
\includegraphics[width=\linewidth]{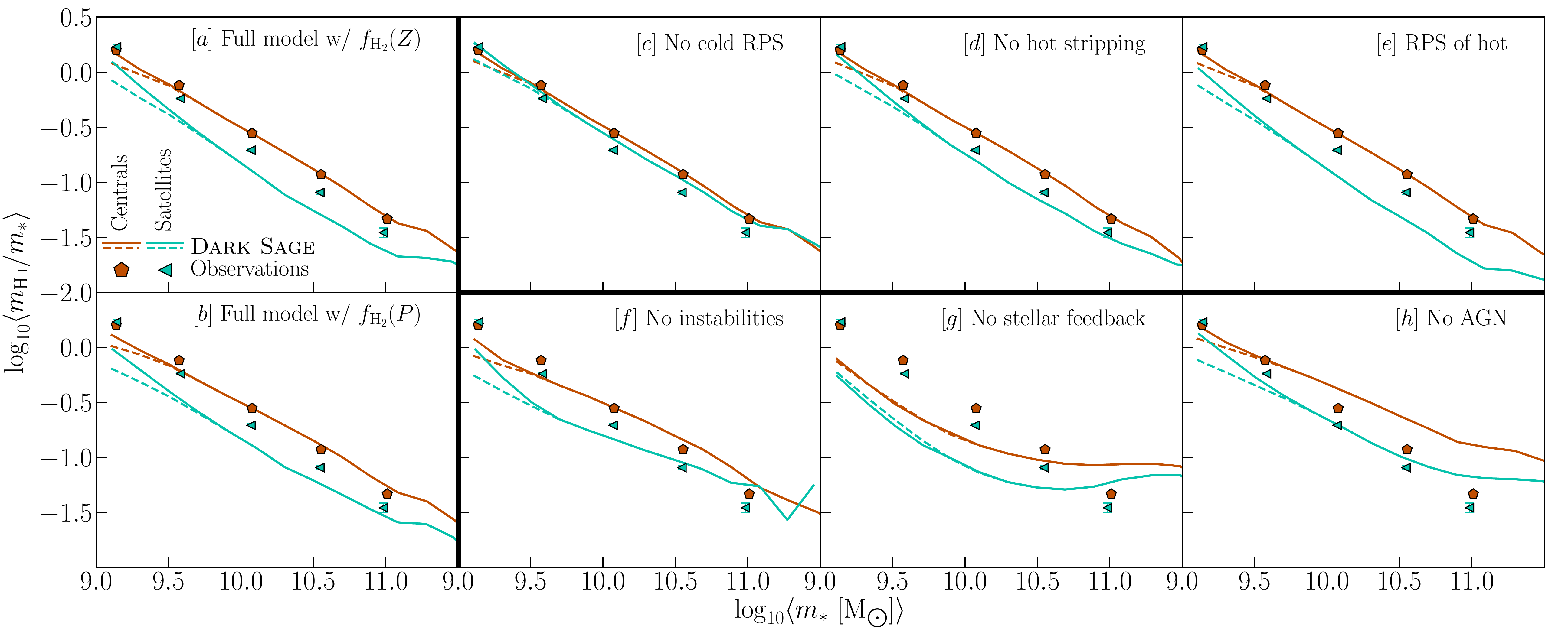}
\caption{Mean atomic hydrogen gas fraction of galaxies as function of mean stellar mass (for pre-determined stellar mass bins) at $z\!=\!0$, split into satellites and centrals.  Each panel includes the same observational data, and shows a different version of {\sc Dark Sage}, with a variation of one process in each case as labelled.  Error on the mean from jackknifing are given for the observational data (Section \ref{ssec:stacking}).  The left-hand panels represent the complete model with the relative \HI and H$_2$ content of each galaxy annulus calculated via a metallicity- or pressure-based prescription ($f_{\rm H_2}(Z)$ and $f_{\rm H_2}(P)$, respectively; see Section \ref{ssec:hih2}).  The three upper right panels consider variations on environmental processes that directly affect satellites.  These assume the metallicity-based prescription for calculating $f_{\rm H_2}$.  The three lower right panels consider variations on evolutionary processes of galaxies and assume the pressure-based prescription for $f_{\rm H_2}$.  Dashed curves consider {\sc Dark Sage} galaxies for all (sub)haloes in the Millennium simulation merger trees (which are composed of at least 20 particles), whereas solid curves only consider (sub)haloes that included 100 or more particles at some point in their history.}
\label{fig:satcen}
\end{figure*}

\begin{figure*}
\includegraphics[width=\linewidth]{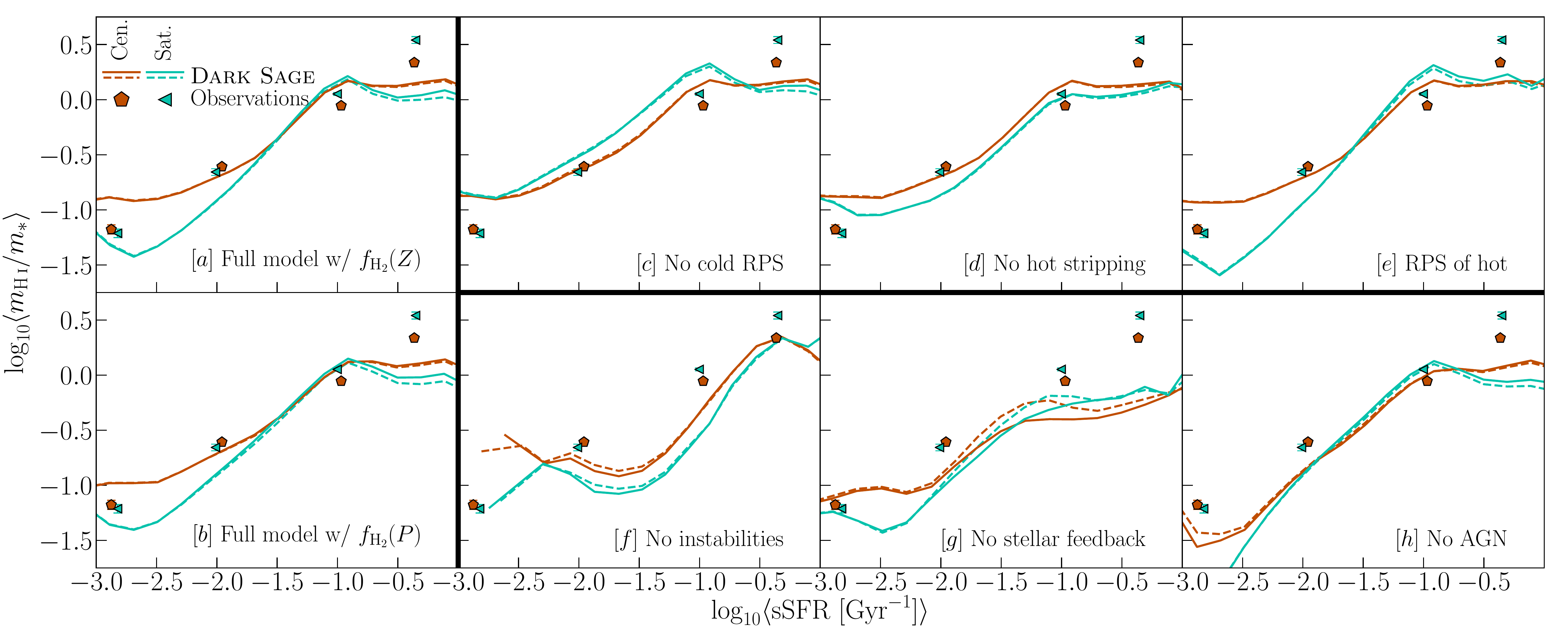}
\caption{As for Fig.~\ref{fig:satcen}, but now mean \HI fraction as a function of mean specific star formation rate (for pre-determined sSFR bins).}
\label{fig:satcen_ssfr}
\end{figure*}

\begin{figure*}
\includegraphics[width=\linewidth]{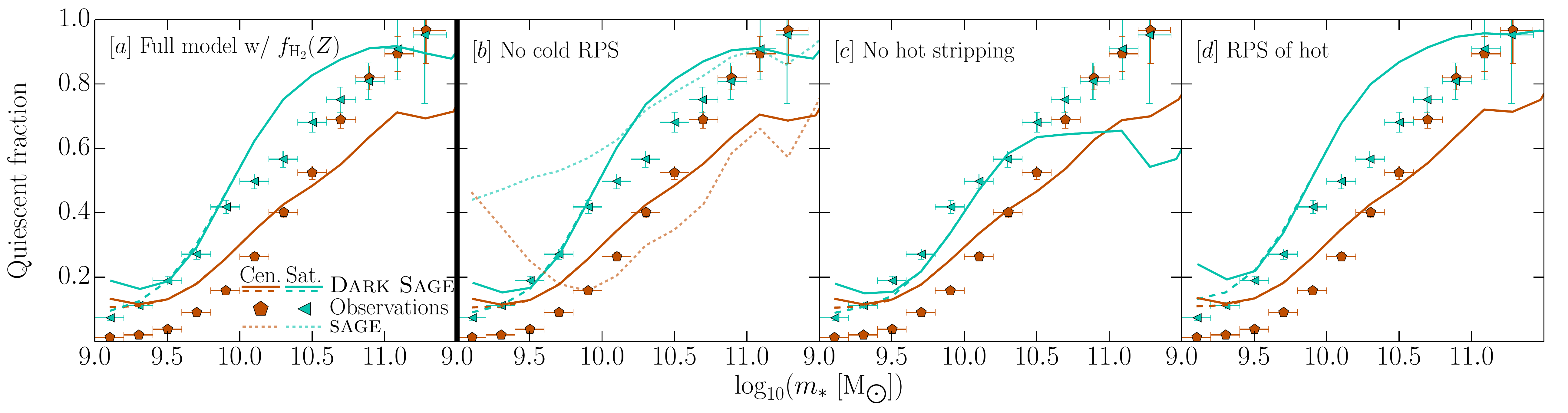}
\caption{Quiescent fraction (${\rm sSFR}\!<\!10^{-11}\,{\rm yr}^{-1}$) of central and satellite galaxies as a function of stellar mass.  Solid curves only account for (sub)haloes with a minimum historical maximum number of particles of 100, whereas dashed curves include the full sample of {\sc Dark Sage} galaxies.  Error bars on the data are Poissonian in the vertical ($\sqrt{N_{\rm quiescent}} / N_{\rm total}$ for each bin) and indicate the width of the bins in the horizontal.  The precise horizontal position of each data point is the mean stellar mass within that bin.  For comparison, panel $b$ also includes the quiescent fraction of galaxies from the original version of {\sc sage} \citep[][dotted curves]{sage}.}
\label{fig:quiescent}
\end{figure*}

In this section, we break both the observed and model galaxies into centrals and satellites, and study the relative differences in their \HI fractions and quiescence.  For the purposes of this paper, a central is the most massive galaxy of a halo, including isolated galaxies that are the sole occupant of a halo.  Satellites constitute all other galaxies.  We perform multiple runs of {\sc Dark Sage} with various environmental and evolutionary processes switched on and off to understand the effect each has on the gas content of galaxies.  In each panel of Fig.~\ref{fig:satcen}, we present the mean \HI fraction of the observed galaxies for fixed values of stellar mass, comparing to one of the runs of {\sc Dark Sage}.  Similarly, in Fig.~\ref{fig:satcen_ssfr}, we assess \HI fraction as a function of specific star formation rate (sSFR).  It is important that we consider this, as \citet{brown15} showed that near-ultraviolet$-r$ band colour, which is a proxy for sSFR, is more tightly related to \HI fraction than stellar mass is.  For {\sc Dark Sage}, we use bins of width 0.2 dex for Figs.~\ref{fig:satcen}--\ref{fig:satenv}.  We plot the log of the mean value along each axis for each bin, for both observed and model galaxies.

In drawing conclusions from comparisons of the \HI fractions of observed and model galaxies, one needs to be mindful of how the relative \HI and H$_2$ content in the model galaxies is determined.  Ideally, we would simultaneously examine the results of {\sc Dark Sage} with the mean H$_2$ fractions of observed galaxies.  Unfortunately, only limited samples of galaxies have their H$_2$ content inferred from CO observations \citep*[e.g.][]{young95,leroy09,saintonge11,boselli14a}, and no H$_2$-blind survey analogous to ALFALFA currently exists (although ongoing surveys such as xCOLDGASS will help alleviate this -- Saintonge et al. in preparation).  There is, however, a wealth of evidence, drawn primarily from these surveys, that closely connects the H$_2$ in these galaxies to their star formation activity \citep[e.g.][]{kennicutt98,rownd99,bigiel08,saintonge11b,saintonge16,boselli14b}.  We can, therefore, use the quiescent fraction of galaxies in observations and models as a first order indication of the relative levels of H$_2$.  We present the quiescent fractions of central and satellite galaxies for the various runs of {\sc Dark Sage}, as a function of stellar mass, alongside the same quantity for our observed sample, in Fig.~\ref{fig:quiescent}.  For our purposes, we consider a quiescent galaxy to be one with ${\rm sSFR}\!<\!10^{-11}\,{\rm yr}^{-1}$.

\subsection{Observations and the full model}
\label{ssec:satcen_full}
As seen in all panels of Figs.~\ref{fig:satcen} \& \ref{fig:satcen_ssfr}, the difference in the mean \HI fraction between the observed centrals and satellites, whether at fixed stellar mass or specific star formation rate, is $\lesssim$\,0.2\,dex.  Both at $m_* \! \lesssim \! 10^{9.5}\,{\rm M}_{\odot}$ and ${\rm sSFR} \! \lesssim \! 10^{-1.5}\,{\rm Gyr}^{-1}$, there is little to separate centrals from satellites.  At higher stellar masses, satellites have less \HI than centrals, suggesting that a significant population of satellites are subject to phenomena (stripping, strangulation, etc.) that deplete their gas reservoirs \citep[\'{a} la][]{gunn72,larson80}.

\begin{figure}
\includegraphics[width=\linewidth]{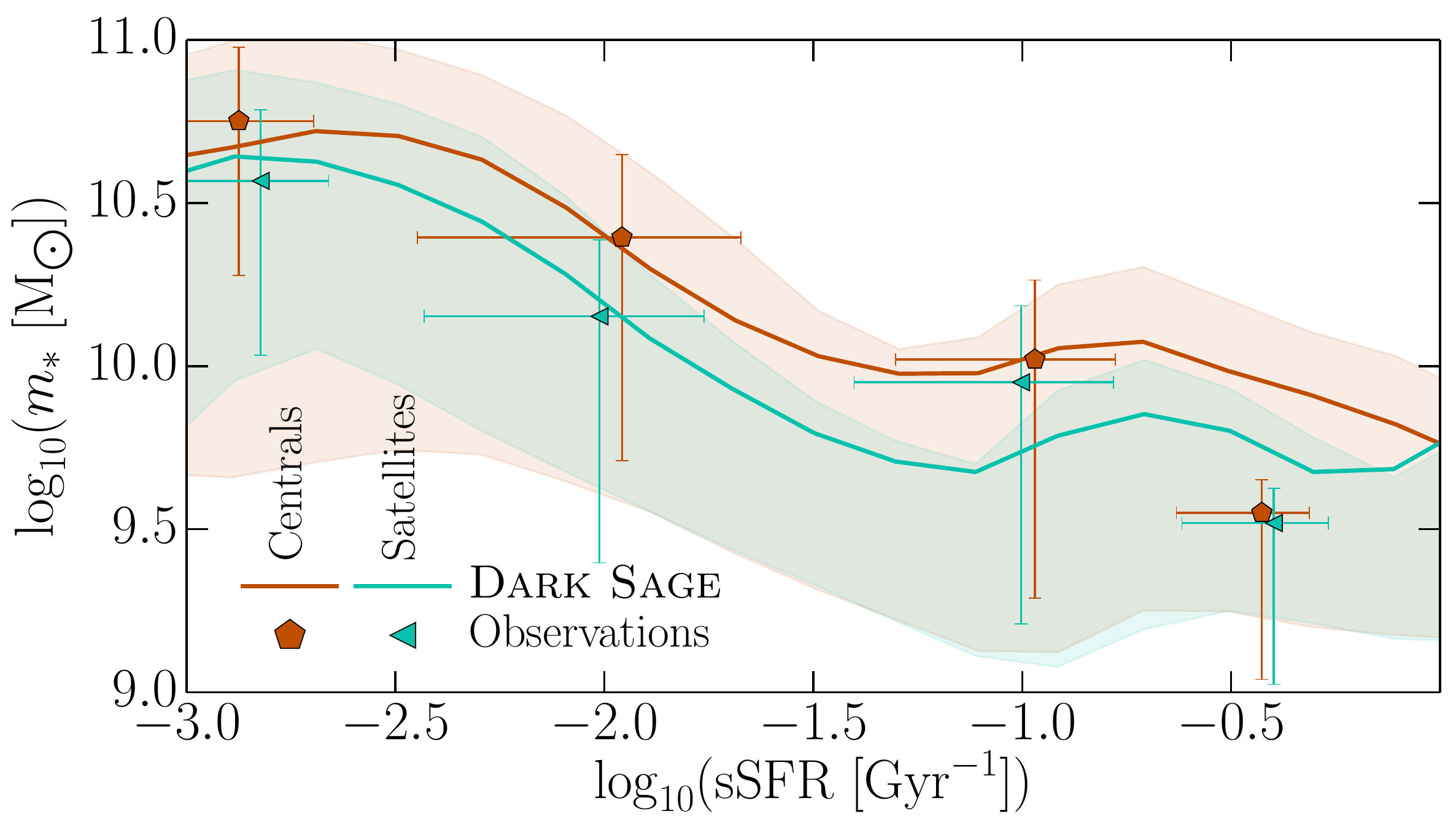}
\caption{Stellar mass of observed and full-model {\sc Dark Sage} galaxies for bins of specific star formation rate, split into centrals and satellites.  Curves and points give the means for the model and observations, respectively, while the error bars and shaded regions cover the 16$^{\rm th}$--84$^{\rm th}$ percentiles (for both variables for the observations).  Only {\sc Dark Sage} galaxies in (sub)haloes that were composed of 100 or more particles at some time in their history are included here.}
\label{fig:ssfr_m}
\end{figure}

At high sSFR, satellites have greater \HI fractions than centrals.  Two effects are important for this comparison.  First is the rate at which gas depletion takes place in satellites.  If depletion processes are slow-acting, the sSFR and \HI fraction will decrease for an individual galaxy at comparable rates.  If they are of moderate speed or fast-acting, the \HI fraction will be observed to decrease before sSFR \citep*[see][]{balogh00,mccarthy08,vollmer12,brown17}.  One might na\"{i}vely except then that at fixed sSFR, satellites would have lower \HI fraction than centrals, which would oppose what is observed.  This expectation would only be legitimate if one were comparing equivalent populations of satellites and centrals (i.e.~of the same stellar mass in each bin), however.  In Fig.~\ref{fig:ssfr_m}, we present the mean stellar mass of satellites and centrals for equivalent bins in sSFR as used in Fig.~\ref{fig:satcen_ssfr}.  Here, we see that at fixed sSFR, centrals tend to have greater stellar mass.  As such, they are biased towards lower gas fractions (cf.~Fig.~\ref{fig:satcen}).

The complete model of {\sc Dark Sage} (regardless of the prescription for the breakdown of atomic and molecular hydrogen) shows a much greater separation in the \HI fractions of satellites and centrals, typically $\lesssim$\,0.65\,dex (panels $a$ \& $b$ of Figs.~\ref{fig:satcen} \& \ref{fig:satcen_ssfr}).  Even though the model is tuned to meet the average \HI fraction of \emph{all} galaxies, this large separation is possible, as the satellites only make up a minority of the galaxies (see below), and because the contribution of the more-massive centrals dominates the average.  As such, {\sc Dark Sage} is in good agreement with observations for the \HI fraction as a function of stellar mass for the centrals (only). This is largely independent of the choice of $f_{\rm H_2}$ prescription, although there is slightly better agreement with observations at the low-mass end when we employ the metallicity-based prescription (cf.~panels $a$ and $b$ of Fig.~\ref{fig:satcen}).  In effect, the properties of the satellites in the model are unconstrained.  \emph{All} results we present concerning satellites are hence predictions (or `\emph{post}dictions') of the model.

38.2 per cent of galaxies in the observational sample are classed as satellites, whereas the {\sc Dark Sage} Millennium sample has 29.7 per cent satellites.  Some difference in these values is expected, due to the nature of group finding versus subhalo finding for the observed galaxies and simulated data, respectively.  In principle, subhalo finding for simulations is a more precise means of identifying satellites \citep[although there is certainly some variation amongst codes -- see][]{onions12,knebe13b,behroozi15}.  Projection effects make it possible for true centrals to be observationally classified as satellites with a group finder, but the converse is unlikely \citep[see, e.g.,][]{campbell15}.  We return to this important point and its potential influence on our results in Section \ref{sec:discussion}.

The {\sc Dark Sage} galaxies also show a flattening in their mean \HI fraction at ${\rm sSFR} \! \gtrsim \! 10^{-1}\,{\rm Gyr}^{-1}$ (panels $a$ and $b$ of Fig.~\ref{fig:satcen_ssfr}), whereas the \HI fractions continue to increase in the observations at high sSFR, with the separation between satellites and centrals also increasing.  A much smaller fraction of the model galaxies occupies these high sSFRs versus the observations, implying galaxies in the model form their first generation of stars too early, leaving them less star-forming at $z\!=\!0$ for an equivalent stellar mass.  Indeed, fig.~A7 of \citet{stevens16} demonstrates that {\sc Dark Sage} galaxies are too star-forming at high redshift.  Similar behaviour is exhibited in Fig.~\ref{fig:ssfr_m}, where the mean stellar mass of {\sc Dark Sage} galaxies as a function of sSFR begins to flatten at $\sim$$10^{-1}\,{\rm Gyr}^{-1}$, whereas the observed galaxies continue towards lower stellar masses.  In the observations, satellites are also found to generally be of lower stellar mass for fixed sSFR versus centrals (Fig.~\ref{fig:ssfr_m}), highlighting why satellites have higher gas fractions at high sSFR (Fig.~\ref{fig:satcen_ssfr}); at low stellar masses, the gradient of mean gas fractions in satellites is steep (Fig.~\ref{fig:satcen}).

To close the loop on the connection between stellar mass, \HI fraction, and specific star formation rate, we compare the quiescent fraction of the full-model {\sc Dark Sage} galaxies against the observations in Fig.~\ref{fig:quiescent}$a$.  Matching the observed quiescent fraction of galaxies as a function of stellar mass has proven challenging for semi-analytic models, especially for satellites \citep[see, e.g.,][]{font08,guo11,guo13,guo16,sage,luo16,henriques17}.  For the centrals, {\sc Dark Sage} exhibits the common feature of overproducing low-mass quenched galaxies and underproducing high-mass quenched galaxies.  For satellite galaxies, at least at low masses, the results are more promising.  Only for $10^{10} \! \lesssim \! m_*/{\rm M}_{\odot} \! \lesssim \! 10^{11}$ is there an overabundance of quiescent satellites. Overall, the quiescent fraction of satellites is notably improved from recent results from both semi-analytic models and hydrodynamic simulations \citep[cf.][]{guo16}. 

\subsection{Environmental processes}
In panels $c$--$e$ of Figs.~\ref{fig:satcen} \& \ref{fig:satcen_ssfr}, we present the mean \HI fraction of {\sc Dark Sage} galaxies after altering the prescriptions of gas stripping in subhaloes in three ways.  The results in panel $c$ of these figures exclude ram-pressure stripping of cold gas entirely (but maintain all other aspects of the model, including hot-gas stripping).  Conversely, those in panel $d$ exclude any consideration of hot-gas stripping (but maintain cold-gas stripping).  Panel $e$ includes both forms of stripping, but uses the alternative ram-pressure prescription for hot gas (Section \ref{ssec:hot}).  Similarly, panels $b$--$d$ of Fig.~\ref{fig:quiescent} present the quiescent fraction of galaxies for these model variants.  In principle, despite these stripping processes only {\it directly} affecting satellites, the hierarchical nature of galaxy formation means that centrals could be causally affected as well.  We do not find any note-worthy changes to the $z\!=\!0$ population of {\sc Dark Sage} central galaxies here, however.

\subsubsection{The overall effect of cold-gas stripping}
\label{ssec:satcen_cold}
In Fig.~\ref{fig:satcen}$c$, we see that, for {\sc Dark Sage} galaxies, without cold-gas stripping, there is little to separate the \HI fraction of satellites and centrals as a function of stellar mass, highlighting that this process is far more effective at reducing \HI fractions of satellites in the model than removal of hot gas.  That said, at least for $m_* \! \lesssim \! 10^{10.2}\,{\rm M}_{\odot}$, the satellite \HI fractions are closer to the observational data than the complete, default model (cf.~panels $a$ and $c$ of Fig.~\ref{fig:satcen}).

In Fig.~\ref{fig:satcen_ssfr}$c$, we see that without cold-gas stripping, the model galaxies now display similar qualitative behaviour to the observed galaxies, in terms of the \HI fraction as a function of sSFR.  That is, at low sSFR, the {\sc Dark Sage} satellites and centrals show the same \HI fraction on average, but for increasing sSFR, the satellites begin to show higher \HI fractions than the centrals.  Although, with this run of the model, this splitting occurs at $\sim$1\,dex lower sSFR than the observations.  Without any fast-acting processes to remove gas from satellites, there are two reasons for satellites showing higher gas fractions than centrals in Fig.~\ref{fig:satcen_ssfr}$c$.  First, the bias of satellites having lower stellar mass than centrals for fixed sSFR, as shown in Fig.~\ref{fig:ssfr_m}, is still present in this run.  Second, even if we were to consider only satellites and centrals of the same stellar mass for a given sSFR, the satellite would have a suppressed supply of fresh gas due to the stripping of hot gas from its subhalo.  The satellite would then have to rely on its instantaneous cold-gas reservoir more heavily for the formation of stars, and hence would need to have a greater mass of \HI at an instant than an equivalent central.

Our results suggest the direct stripping of cold gas in galaxies should have a large impact on their \HI content (which we reaffirm in Section \ref{sec:satenv}), but that the current implementation in {\sc Dark Sage} is systematically too efficient at removing \HInospace.  Potential solutions could include: (i) non-instantaneous stripping of the cold gas within the annuli, (ii) a more physically strict criterion for when cold gas stripping is allowed (rather than the current case that $m_{\rm cold} \! + \! m_* \! > \! m_{\rm hot}$), (iii) adding inclination effects to equation (\ref{eq:rps}), or (iv) the observed \HI masses include gas that the model would not consider `part of the galaxy' (which is hard to define, even when one has full three-dimensional information like in a hydrodynamic simulation -- see \citealt{stevens14}).  However, each of these is unlikely to be satisfactory by itself, as this will have consequences for the relative \HI fractions of satellites in different environments (as the strength of stripping is dependent on local environment), which, as we show in Section \ref{sec:satenv}, is an area the model performs well in.  We discuss this further in Section \ref{sec:discussion}.

Where cold-gas stripping has a negligible effect is on the quiescent fraction of galaxies (cf.~panels $a$ and $c$ of Fig.~\ref{fig:quiescent}).  As indicated by equation (\ref{eq:rps}), ram-pressure stripping of cold gas is more dominant in the low-surface-density areas of satellites' discs, towards their outskirts.  The hydrogen in annuli with low $\Sigma_{\rm gas}$ will also be predominantly atomic (regardless of the specific choice of $f_{\rm H_2}$ prescription -- see Section \ref{ssec:hih2} of this paper and section 3.4 of \citealt{stevens16}).  Star-forming gas is predominantly molecular and hence deeper in the potential well.  As a result, star-forming gas is less susceptible to ram-pressure stripping, and thus this process has little impact on the quiescent fraction of satellites.

We note that even without cold-gas stripping, our quiescent-fraction results are a significant improvement over {\sc sage} \citep[][from which {\sc Dark Sage} was developed]{sage} for both centrals and satellites, especially at low masses.  To highlight this, we have included results from {\sc sage} in Fig.~\ref{fig:quiescent}$b$ (shown in panel $b$ as {\sc sage} does not include cold-gas stripping).

\subsubsection{The overall effect of hot-gas stripping}
\label{ssec:satcen_hot}
Fig.~\ref{fig:satcen}$d$ shows that when there is no given environmental mechanism to remove the hot-gas reservoir of subhaloes, the mean \HI fraction of {\sc Dark Sage} galaxies is remarkably close to observations for a given stellar mass.  The maintenance of that reservoir allows gas to continue to cool onto satellite galaxies in the model for longer.  Based on the severity of the difference seen in the model when cold-gas stripping is removed, it is safe to say the greatest reason removing hot-gas stripping from the model has an effect on the \HI fraction of galaxies is because hot gas shields the cold gas from being stripped (see Section \ref{ssec:rps}).  This minimises the difference between the gas fractions of satellites and centrals of the same mass, which is favourable next to observational data.  

Excluding hot-gas stripping does, however, lead to satellites having systematically lower \HI fractions than centrals as a function of sSFR, for all sSFR (Fig.~\ref{fig:satcen_ssfr}$d$).  This is at odds with observations.  Without hot-gas stripping, satellites can continue to accrete plenty of fresh gas, which, by design of the model, settles onto the disc in an approximately exponential form.  Because cold-gas stripping is still present, the outer gas that cools is quickly removed, whereas the inner gas enhances star formation.  Hot-gas stripping therefore affects the molecular-to-atomic gas ratio of satellites, where, without it, galaxies would have higher sSFR at fixed \HI fraction, as seen in Fig.~\ref{fig:satcen_ssfr}$d$.

The impact of hot-gas stripping on the star-forming gas content of satellites is highlighted further in Fig.~\ref{fig:quiescent}$d$.  Without hot-gas stripping, the fraction of quiescent galaxies drops dramatically for satellites with $m_* \! \gtrsim 10^{10}\,{\rm M}_{\odot}$.  This result would be unsurprising for any semi-analytic model \citep[see, e.g., the difference in gradual and instantaneous hot-gas stripping from][]{font08}, but the point here is that even when cold-gas stripping is included (which is not the case in many models), hot-gas stripping remains the dominant process for regulating the quiescent fraction of satellites.  Moreover, we see that low-mass galaxies in {\sc Dark Sage} are not plagued by being over-quenched from hot-gas stripping, which has been a persistent issue with semi-analytic models in the past, as raised in Section \ref{ssec:satcen_full}.

Finally, in Figs.~\ref{fig:satcen}$e$, \ref{fig:satcen_ssfr}$e$, and \ref{fig:quiescent}$e$, we show the results of {\sc Dark Sage} with the alternative, ram-pressure-like prescription for hot gas (Section \ref{ssec:hot}).  With this implementation, high-sSFR satellites begin to exhibit slightly higher \HI fractions than their central counter-parts, qualitatively more in line with the observations, whereas the opposite is true for high-mass galaxies.  There is also a minimal increase in the fraction of quiescent satellites.  By and large though, the two implementations of hot-gas stripping are relatively consistent when it comes to the mean differences in gas fractions of satellites and centrals.

\subsection{Evolutionary processes}
\label{ssec:evolution}
In this subsection, we assess processes that affect all galaxies (regardless of whether they are centrals or satellites).  We focus on the effects of (i) disc instabilities, (ii) stellar feedback and evolution, and (iii) AGN feedback, which all appear to play some role in regulating the relations between \HI fraction, stellar mass, and sSFR.  We present results from {\sc Dark Sage} with each of these features independently switched off in panels $f$--$h$ of Figs.~\ref{fig:satcen} \& \ref{fig:satcen_ssfr}.  

It should be noted that unlike for the previous runs where we altered stripping processes, simply removing feedback or instabilities from the model has a significant impact on mass functions and scaling relations, which the model is usually constrained to meet.  These runs have not been recalibrated; they are, therefore, an attempt to gauge what the \HI fractions of galaxies would be if each physical phenomenon did not exist, rather than attempting to recover realistic galaxies in the absence of said phenomena.  This provides a more direct measure of how each process independently helps to shape the \HI fractions of galaxies.  

We also note that, despite using the metallicity-based prescription for determining the fraction of gas in each annulus in the form of \HI and H$_2$ (equation \ref{eq:RH2}) as the new default for {\sc Dark Sage}, we have assumed the pressure-based prescription for the runs without feedback or instabilities.  As noted in Section \ref{ssec:hih2}, passive star formation from H$_2$ will only ignite after initial metal enrichment, which happens when a galaxy first experiences a starburst from either an instability or merger.  Without instabilities in the model then, only galaxies that experience a merger will have the passive H$_2$ star formation channel open to them.  This channel of star formation is the dominant mechanism by which stars are formed in low-mass galaxies.  Because the majority of galaxies in the Millennium simulation merger trees do not have a recorded merger, one would end up with a non-physically large number of star-less galaxies with large gas reservoirs.  A similar problem arises when stellar evolution is removed from the model, as this removes metal enrichment entirely.  Both cases are resolved by switching to the mid-plane pressure prescription for $f_{\rm H_2}$, which is unaffected by gas metallicity.  For consistency, we also use this prescription for the run without AGN feedback (but this does not make a significant difference in this case).

\subsubsection{The effect of disc instabilities}

As shown in Fig.~\ref{fig:sfchannels}, most star formation in low-mass {\sc Dark Sage} systems comes from the passive H$_2$ channel, whereas the high-mass systems formed most of their stars through the instability channel.  By eliminating any consideration of disc instabilities from the model, star formation, that would have otherwise more directly harnessed available \HI reservoirs, becomes progressively less efficient towards the high-mass end.  As a result, come $z\!=\!0$, not only are there fewer galaxies of high stellar mass, but the gradient of the mean \HI fraction as a function of stellar mass has become shallower, which we show in Fig.~\ref{fig:satcen}$f$.

With less integrated star formation having occurred in the lead up to $z\!=\!0$ without instabilities in the model, galaxies are a lot more gas rich on average.  Also, because instabilities are unresolved, most of the gas remains gravitationally unstable in dense clumps, and hence there is more H$_2$ than \HInospace.  This means that the (instantaneous) star formation rates are actually higher at $z\!=\!0$ than they were for the full model.  This is seen in Fig.~\ref{fig:satcen_ssfr}$f$, where the average \HI fraction for a given sSFR has decreased.

\citet{stevens16} showed that instabilities regulated the stellar specific angular momentum of {\sc Dark Sage} galaxies as a function of stellar mass.  It is no coincidence that they also regulate the gas fractions of galaxies; recent analytic work by \citet{ob16} and results from hydrodynamic simulations \citep{lagos17} have shown that the gas fraction and specific angular momentum of a galaxy are inherently related.  We will address this connection in more specific detail with {\sc Dark Sage} in future work.

\subsubsection{The effect of stellar feedback and evolution}
Without stellar feedback or stellar evolution, an equivalent amount of gas can be used to form a greater mass of stars.  This is because the instantaneous recycling approximation no longer needs to be upheld, nor does a fraction of the cold gas in a star-forming annulus need to be reserved for being reheated out of the disc.  Hence, the run of {\sc Dark Sage} without stellar feedback (or evolution) shows a deficit in the mean \HI fraction for fixed sSFR, as in Fig.~\ref{fig:satcen_ssfr}$g$.  

Provided galaxies continue to accrete fresh gas, the lack of gas reheating raises the frequency of disc instabilities.  There also tends to be greater cold-gas reservoirs in galaxies during mergers.  These both not only mean that starbursts become stronger in the model, but also that black holes are able to accrete gas faster.  Larger black holes lead to stronger AGN feedback, which shuts down fresh gas accretion at lower redshift.  This leads to a deficit in the stellar mass function in the range $10^{10.5} \! \lesssim \! m_*/{\rm M}_{\odot} \! \lesssim \! 10^{11.5}$ and an excess of galaxies with lower masses.  Resultantly, at $z\!=\!0$, the gas fractions of galaxies at lower masses are systematically reduced, as seen in Fig.~\ref{fig:satcen}$g$.  In this situation, only galaxies that experience many mergers can accumulate enough cold baryons to grow to larger stellar masses.  Because all mergers are more gas rich, the highest-mass galaxies have larger \HI fractions than in the full model.  This is also seen in Fig.~\ref{fig:satcen}$g$.

\subsubsection{The effect of active galactic nuclei}
Without an AGN engine to regulate the rate at which galaxies accrete gas, all galaxies, but especially the high-mass ones, become extra gas-rich.  This is seen in Fig.~\ref{fig:satcen}$h$.  Where an AGN makes a smaller difference in {\sc Dark Sage} is with how gas fraction varies with sSFR.  Fig.~\ref{fig:satcen_ssfr}$h$ shows that the most star-forming galaxies have almost unchanged \HI fractions when AGNs are removed from the picture.  On the other hand, the more quiescent galaxies have lower \HI fractions for fixed sSFR in this case.  This comes back to the extra accretion galaxies receive; much of this extra gas can quickly be consumed in star formation, leading to higher sSFRs (which are averaged over a finite time-scale) for a given instantaneous gas fraction.  This is equivalent to getting lower instantaneous gas fractions for a given sSFR.

\section{Satellites in various environments}
\label{sec:satenv}

In this section, we investigate how the \HI fractions of satellite galaxies vary as a function of parent halo mass, which serves as a measure of environment.  We consider three bins of halo mass: (i) $M_{\rm vir} \! < \! 10^{12}\,{\rm M}_{\odot}$, (ii) $10^{12} \! \leq \! M_{\rm vir}/{\rm M}_{\odot} \! < \! 10^{13.5}$, and (iii) $10^{13.5} \! \leq \! M_{\rm vir}/{\rm M}_{\odot} \! < \! 10^{15}$.  For the {\sc Dark Sage} galaxies, virial masses of haloes are directly given from the Millennium simulation.  As discussed in Section \ref{sec:obs}, the halo mass measurements for the observed galaxies follow the group-finding algorithm of \citet{yang07}.  

There is an argument to be made that, for a fairer comparison, one should run the same group-finding algorithm on the {\sc Dark Sage} output as was used for the observations, and use those returned halo masses.  We have tested that by rank ordering the stellar and virial masses of haloes, and reassigning halo masses by the ranked virial mass, any changes to our results are entirely negligible.  Of course, this only mimics part of the method used in estimating the halo masses of the observed galaxies.  Any further differences would come as a result of cross-contamination of centrals and satellite populations (as raised in Section \ref{ssec:satcen_full}).  We have opted for the simpler, more digestible comparison for our main results (presenting the true model output with the observational data as they are), but we discuss the effect cross-contamination could have in Section \ref{sec:discussion}.

\begin{figure*}
\includegraphics[width=\linewidth]{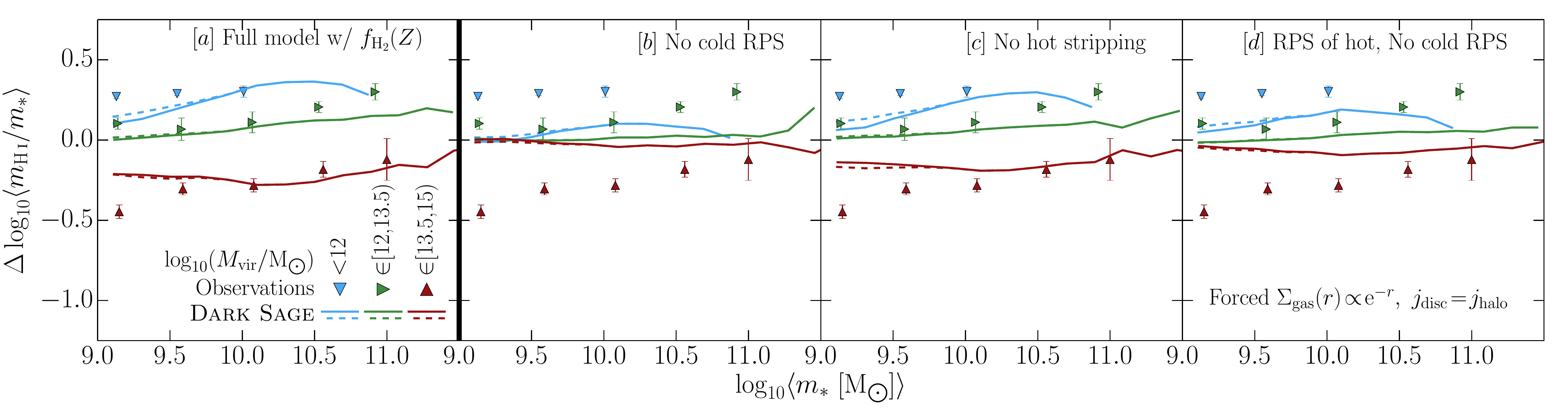}
\includegraphics[width=\linewidth]{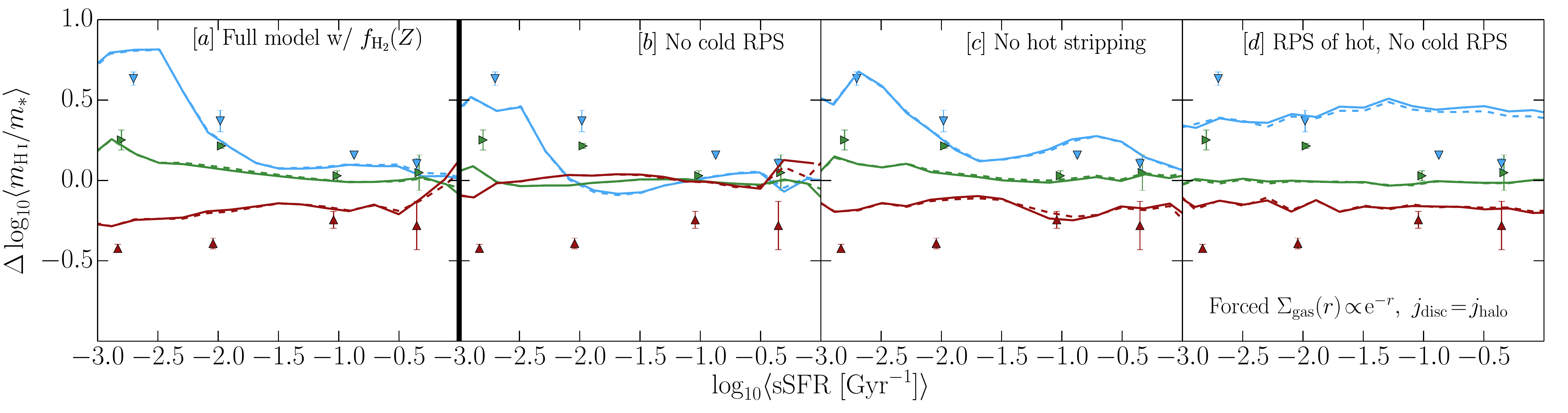}
\caption{Difference in mean atomic hydrogen gas fraction, as a function of stellar mass (top row) and specific star formation rate (bottom row), for \emph{satellite} galaxies at $z\!=\!0$ for bins of parent halo mass.  Each panel right of the thick vertical line changes an environmental process in {\sc Dark Sage}, as labelled.  Panel $d$ further recalculates \HI fractions in post-processing by redistributing matter into a new, exponential profile (see text for details).  Solid curves only consider (sub)haloes composed of at least 100 particles at some point in their history, whereas dashed curves include the full sample of {\sc Dark Sage} galaxies.  All panels include the same observational data, with errors on the means from jackknifing.}
\label{fig:satenv}
\end{figure*}

\begin{figure*}
\includegraphics[width=\linewidth]{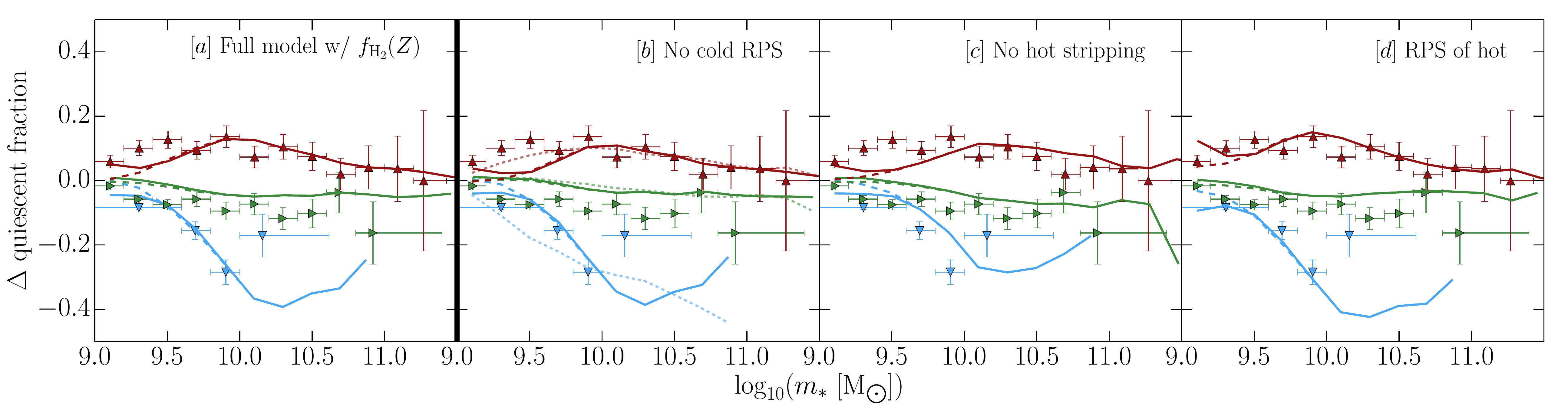}
\caption{Difference in the quiescent fraction of satellite galaxies within halo mass bins from the overall quiescent fraction of satellites.  This is given as a function of stellar mass for three halo mass bins (see the legend of Fig.~\ref{fig:satenv}).  Error bars on the data are Poissonian in the vertical and indicate the full width of the bins in the horizontal.  The precise horizontal position of each data point is the mean stellar mass within that bin.  Dotted curves in panel $b$ compare results from the original version of {\sc sage} \citep{sage}.}
\label{fig:quiescent_env}
\end{figure*}

The main results of this section are presented in Fig.~\ref{fig:satenv} \& \ref{fig:quiescent_env}.  In Fig.~\ref{fig:satenv}, instead of presenting the mean \HI fractions of satellite galaxies for the various halo mass bins on the $y$-axis (e.g., as in \citealt{brown17}), we present $\Delta \log_{10}(m_{\mathrm{H}\,\textsc{i}} / m_*)$, which is the \emph{separation} between the mean \HI fraction curve for that halo mass bin and the mean curve for all satellite galaxies (the curves and points in Figs.~\ref{fig:satcen} \& \ref{fig:satcen_ssfr}).  This means we can directly compare how environment splits the \HI content of satellite galaxies in the observations and model without being concerned about any normalisation issues between the two (which have been discussed in Section \ref{sec:satcen}).  Similarly, in Fig.~\ref{fig:quiescent_env}, rather than presenting the raw quiescent fraction of satellites in different environments, we show the \emph{difference} in those fractions from the \emph{overall} quiescent fraction of satellites (given in Fig.~\ref{fig:quiescent}).

\subsection{Full model versus observations}
In Fig.~\ref{fig:satenv}, we show how halo mass (environment) affects the \HI fraction of satellite galaxies in the full {\sc Dark Sage} model and the observations with stellar mass (panel $a$) and sSFR (panel $e$).  For $m_* \! \gtrsim \! 10^{9.5}\,{\rm M}_{\odot}$, the observations and model are in remarkable agreement: better than 0.1\,dex for the mean \HI fraction in all halo mass bins.  Only at lower stellar masses do we begin to see a discrepancy, where the model shows less environmental splitting than the observations.  This is true regardless of whether we only consider subhaloes from the $z\!=\!0$ snapshot of Millennium with a historical maximum number of particles of $N_{\rm p,max} \! \geq \! 100$, or use all subhaloes with $\geq$20 particles.  In Fig.~\ref{fig:quiescent_env}$a$, we show that {\sc Dark Sage} also recovers the \emph{relative} role environment plays on the quiescent fraction of satellites to an encouraging degree.  In this case, results at low-$m_*$ are more in line with observations when we apply the cut of $N_{\rm p,max} \! \geq \! 100$.

An important consideration is whether the distribution of satellites in the various halo mass bins is consistent between the observations and model data.  While environmental processes will not inherently care about the number of galaxies they act upon, the differences in \HI fractions and quiescent fraction of satellites in a given environment from the overall mean will depend on the contribution of satellites in that environment \emph{to} the overall mean.  As such, we list the number of satellites in each halo mass bin for the observed and model samples in Table \ref{tab:Nsat}.

\begin{table*}
\centering
\begin{tabular}{l r r r} \hline
 & \multicolumn{3}{c}{$N_{\rm sat}$}\\
$\log_{10}(M_{\rm vir}/{\rm M}_{\odot})$ & $<\,12$ & $\in [12,13.5)$ & $\in [13.5,15)$\\\hline
Observations & 1\,030 (11.9\%) & 3\,486 (40.2\%) & 4\,148 (47.9\%) \\
{\sc Dark Sage}, all & 194\,240 (13.3\%) & 704\,781 (48.4\%) & 544\,154 (37.4\%) \\
{\sc Dark Sage}, $N_{\rm p,max}\!\geq\!100$ & 155\,430 (12.9\%) & 585\,254 (48.6\%) & 453\,286 (37.6\%) \\\hline
\end{tabular}
\caption{Number of satellite galaxies in the various halo mass bins for both the observations and full model of {\sc Dark Sage} (which, apart from the cut of $m_* \! \geq \! 10^9\,{\rm M}_{\odot}$, comes from Millennium), for which results are shown in Figs.~\ref{fig:satenv} \& \ref{fig:quiescent_env}.  $N_{\rm p,max}$ refers to the historical maximum number of particles that each subhalo was composed of in the Millennium merger trees.}
\label{tab:Nsat}
\end{table*}

Another important consistency check between the observations and models is the probability distribution of halo masses (within each halo mass bin).  Because the group finder used to associate halo masses with observed galaxies drew from a halo mass function produced by $N$-body simulations whose cosmology is very similar to Millennium's \citep[cf.][]{millennium,warren06}, we should expect these to be consistent.  However, because the observed sample of galaxies is limited in its volume and stellar mass range, differences in the distributions of halo masses can arise.  We have tried extracting, as precisely as possible, a set of haloes from the Millennium simulation that matches the halo catalogue from the observations + group finder.  While the smaller sample size introduces noise, no significant conclusions from our figures come about by making this selection (not shown here).  This is in spite of the fact that the specific cases of the Coma and Hercules clusters have a large influence on the largest halo mass bin for the observational data.

\subsection{The relative impact of cold-gas stripping}
In Fig.~\ref{fig:satcen}, we demonstrated that cold-gas stripping in {\sc Dark Sage} was almost entirely responsible for the systematic separation in \HI fractions between satellite and central galaxies of the same stellar mass.  Similarly, removing cold-gas stripping from the model almost eliminates any environmentally driven splitting in the \HI fractions of satellite galaxies.  This is seen not only as a function of stellar mass in Fig.~\ref{fig:satenv}$b$, but also as a function of sSFR in Fig.~\ref{fig:satenv}$f$.

The results of {\sc Dark Sage} presented here are in contrast to those of the \citet{gp14} version of the {\sc galform} semi-analytic model presented alongside the same observations in \citet{brown17}.  That is, despite the model being run on the same simulation \citep[Millennium;][]{millennium} and lacking a prescription for ram-pressure stripping of cold gas entirely, {\sc galform} still displayed \HI fraction splitting with halo mass for satellite galaxies.  There are many differences between {\sc Dark Sage} and {\sc galform} in terms of the specifics of how galaxies are evolved \citep[for details on {\sc galform}, see][]{cole00,benson10,lacey16}.  One vital aspect, which we consider here, is how the \HI and \Htwo surface density profiles of discs are ultimately derived.  We note that a modified set of merger trees \citep{jiang14} for Millennium is employed for {\sc galform}, but this algorithm is only expected to have a significant impact on semi-analytic galaxy properties for higher-resolution $N$-body simulations \citep[see][]{merson13,jiang14}.

The key feature of {\sc Dark Sage} is that the full two-phase angular-momentum structure of discs (and hence their radial structure also) is numerically evolved self-consistently.  Similar to many other semi-analytic models \citep[cf.][]{hatton03,somerville08,guo11}, {\sc galform} only evolves the \emph{total} angular momentum of discs (gas and stars together), and then assumes that discs always carry an exponential surface density profile with radius.  Although discs are seeded as approximately exponential in {\sc Dark Sage}, they develop strong cusps come $z\!=\!0$, partly caused by instability-induced radial flows of gas \citep{stevens16}.  These cuspy gas discs are qualitatively in agreement with both observations and hydrodynamic simulations \citep[see][respectively]{bigiel12,stevens17}.  A cuspy disc will have higher $R_{\rm H_2}(r)$ at small $r$ and relatively lower $R_{\rm H_2}(r)$ at large $r$ (e.g., see Equation \ref{eq:RH2}).  Hence, a {\sc Dark Sage} galaxy will have more H$_2$ than a {\sc galform} galaxy with the same stellar mass, cold gas mass, and halo properties.  Some processes that affect the \HI content of {\sc galform} galaxies might, therefore, affect the H$_2$ content of {\sc Dark Sage} galaxies as well/instead.  We quantify the impact of this further in Section \ref{ssec:disc_modelling}.

Within {\sc Dark Sage}, cold-gas stripping only affects the star formation activity of low-$m_*$ satellites in the most massive haloes.  This is evidenced by Fig.~\ref{fig:quiescent_env}$b$, where only in this regime is there a notable change to the relative quiescent fraction when cold-gas stripping is removed.  The combination of these low-mass galaxies having shallow local potential wells and the high-mass haloes in which they reside inflicting strong ram pressure means that even the molecular gas in the satellites is susceptible to being directly stripped.  

We also include results from the regular {\sc sage} model \citep{sage} in Fig.~\ref{fig:quiescent_env}$b$.  The fact that {\sc sage} does not include cold-gas stripping, but is still able produce a decent result for the \emph{relative} role of environment on quiescent fraction, is testament to cold-gas stripping only being of secondary importance for satellites' quiescence.  Differences seen between {\sc sage} and {\sc Dark Sage} here are the result of detailing the evolution of discs' structure.

\subsection{The relative impact of hot-gas stripping}
As discussed in Section \ref{ssec:satcen_hot}, any changes to the mean \HI fraction of satellite galaxies from removing hot-gas stripping from {\sc Dark Sage} were mostly caused by a reduction in the opportunity for cold-gas stripping to take place.  As a result, the splitting by environment of satellites' \HI fractions as a function of stellar mass is also reduced from the full model, as shown in Fig.~\ref{fig:satenv}$c$.  Despite the fact that removing hot gas from the model brought the difference in the mean \HI fractions of satellites and centrals more in line with observations (Fig.~\ref{fig:satcen}$c$), the splitting of satellites' \HI fractions by halo mass has become weaker than the observations for this run.  This implies that while stripping processes in {\sc Dark Sage} are systematically too strong (Fig.~\ref{fig:satcen}$a$), they are largely of the correct \emph{relative} strength in haloes of different mass.

Hot-gas stripping plays a major role in regulating the quiescent fraction of satellites, especially at higher stellar masses (Section \ref{ssec:satcen_hot}).  As shown in Fig.~\ref{fig:quiescent_env}$c$, removing hot-gas stripping also reduces the environmental splitting of the quiescent fraction of satellite galaxies with $m_* \! \lesssim \! 10^{10}\,{\rm M}_{\odot}$.  Satellites of equivalent stellar mass (or subhalo mass at infall) experience greater depletion of their hot-gas reservoir when falling into more massive haloes.  This is true regardless of whether we prescribe hot gas to be stripped at the same rate as dark matter or impose ram pressure on the hot gas (cf.~panels $a$ and $d$ of Fig.~\ref{fig:quiescent_env}).  However, Fig.~\ref{fig:quiescent_env} also highlights that a \emph{relative} environmental dependence would still be present (and consistent with observations) for the quiescent fraction of satellites without any stripping processes.  This is because the hot-gas reservoirs of satellites are not allowed to grow through cosmological accretion within {\sc Dark Sage} (we come back to this in Section \ref{sec:discussion}).  As a result, the longer a galaxy exists as a satellite, the less gas it will have available to accrete, and thus the more quiescent it will become.  Satellites in larger haloes have longer orbital and merging time-scales, and therefore will have greater opportunity to consume their gas and become quiescent.

\subsection{The impact of disc structure modelling}
\label{ssec:disc_modelling}

As an experiment in post-processing, we reset the specific angular momentum of the {\sc Dark Sage} discs so that $j_{\rm disc}\!=\!j_{\rm halo}$, and we redistribute both gas and stars in the disc such that they follow $\Sigma_{\rm gas}(r) \! \propto \! {\rm e}^{-r}$ with a common scale radius.  This meets the disc model employed in the standard version of {\sc sage} \citep{sage}.  We then recalculate the \HI and \Htwo masses of each galaxy according to the $f_{\rm H_2}(P)$ prescription \citep[which matches the method of {\sc galform} -- see][]{lagos11}.  We further recalculate their SFR based on their adjusted \Htwo content, assuming $\Sigma_{\rm SFR}(r) \! = \! \epsilon_{\rm SF} \Sigma_{\rm H_2}(r)$, maintaing $\epsilon_{\rm SF} \! = \! 1.3 \! \times \! 10^{-4}\,{\rm Myr}^{-1}$.  This allows us to estimate what differences we would find regarding the relative influence of environment on the \HI fractions of satellites if {\sc Dark Sage} did not include a detailed treatment of the angular momentum of discs and did not allow them to be cuspy.  We present results for this for a run of the model that uses the ram-pressure prescription for hot-gas stripping, and excludes cold-gas stripping entirely.  These choices give us the most directly comparable results to those from {\sc galform} version GP14+GRP \citep{gp14} published in figure 7 of \citet{brown17}.  Our results are given in panels $d$ and $h$ of Fig.~\ref{fig:satenv}.

Fig.~\ref{fig:satenv}$d$ shows that an environmental splitting of \HI fractions as a function of stellar mass caused entirely by hot-gas stripping can be recovered for a more rudimentary model of galaxy discs.  The strength of this splitting is still significantly less than what is observed in real galaxies.  It is also less than half the strength seen in {\sc galform} \citep[cf.][]{brown17}.  Meanwhile, the splitting with sSFR has become flat, now too weak at low sSFR and too strong at high sSFR versus observations (Fig.~\ref{fig:satenv}$h$).  Of course, because the gas redistribution is done in post-processing, the evolution of the galaxies is not affected.  It is an obvious statement that if aspects of galaxy evolution within one model were changed to match that of another model, the two models would end up giving very similar results.  Such an exercise is hence not particularly informative.  What the results of panels $d$ and $h$ of Fig.~\ref{fig:satenv} do tell us is that the method for determining $f_{\rm H_2}$ only has a limited potential contribution to the dominance of cold-gas stripping in driving the environmental dependence of satellites' \HI fractions in {\sc Dark Sage}.  In other words, these results still favour ram-pressure stripping of cold gas to be the main mechanism for curbing the \HI content of satellite galaxies.

\section{Caveats, discussion and conclusions}
\label{sec:discussion}

We have shown that the \emph{relative} role that environment is observed to have on the \HI fractions of satellites \citep{brown17} can be recovered within a semi-analytic model of galaxy formation with a self-consistent evolution of disc structure, namely {\sc Dark Sage} \citep{stevens16}.  However, this model also predicts the mean \HI fractions of satellites as a whole to be too low.  In the current design of the model, these two results are inherently tied together.  One possible explanation is that the model is missing a physical process that would systematically (and exclusively) raise the gas content of satellites, irrespective of their environment.  

One process hypothesised to help explain why low-mass galaxies in denser environments are observed to have brighter X-ray haloes \citep{mulchaey10} is `confinement pressure'.  The idea here is that the temperature differential between the hot gas of a satellite and the hotter intracluster medium (ICM) creates an inward pressure, thus helping that satellite to retain its hot gas and dampening the effect of hot-gas stripping.  However, it has been shown with both a simple analytic model and the results of hydrodynamic simulations that confinement pressure only has a strong dampening effect on hot-gas stripping once a satellite reaches pericentre, at which point most of its hot gas has been removed already \citep{bahe12}.  The satellite galaxies in the {\sc Dark Sage} results without hot-gas stripping presented in this paper were also prescribed to maintain any reheated gas from feedback within their own hot reservoir (Section \ref{ssec:hot}).  In effect, this run of the model shows an extreme for what confinement pressure could achieve.

More recently, using a cosmological hydrodynamic simulation focussed on a galaxy cluster, \citet*{quilis17} have suggested that the hot gas of satellite galaxies can be replenished by accretion from the ICM.  This helps to counter-act the stripping of hot gas, where cold-gas stripping is shown by \citeauthor{quilis17} to have a greater net impact on these galaxies.  Complementary results from the EAGLE simulations have explicitly shown that the gas accretion rates through a fixed 30-kpc aperture around satellite galaxies still show a strong dependence on environment \citep{voort17}.  Note, though, that gas accretion rates through a fixed aperture can be wildly different to accretion rates onto a (sub)halo \citep{stevens14}.

Unlike central galaxies, which accrete hot gas cosmologically, satellites cannot acquire hot gas from an external source within the current {\sc Dark Sage} model.  This feature remains a shared standard for semi-analytic models in general \citep[see the latest versions of other models, e.g.][]{gargiulo15,henriques15,lacey16}.  If subhaloes were able to accrete hot gas from the ICM within a semi-analytic framework, this should systematically raise \HI fractions and lower quiescent fractions.  Both of these outcomes would be favourable for {\sc Dark Sage}, potentially offering a solution (at least partially) for simultaneously recovering the absolute and relative impacts of environmental stripping processes.

\begin{figure*}
\includegraphics[width=0.34\linewidth]{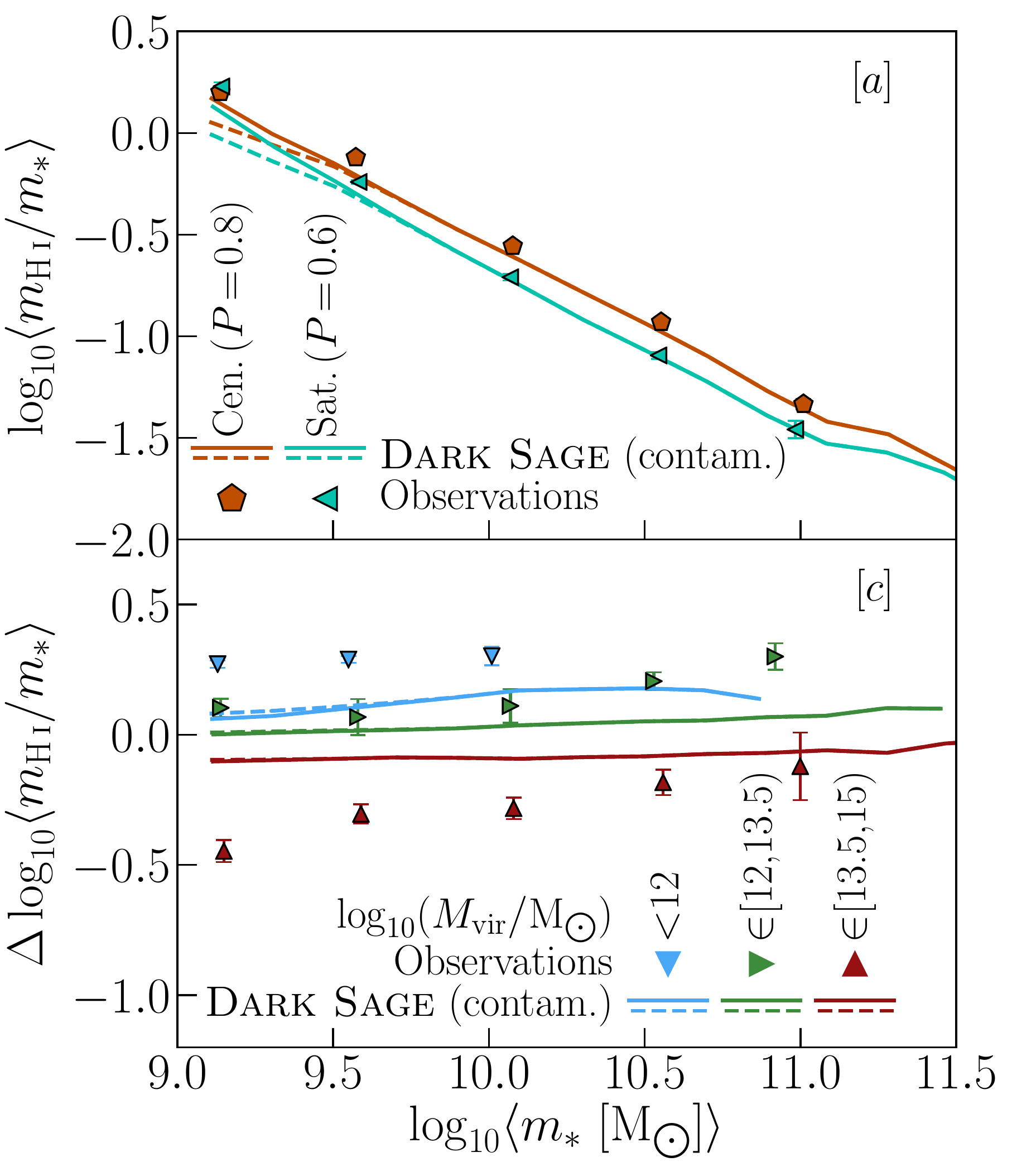}
\includegraphics[width=0.34\linewidth]{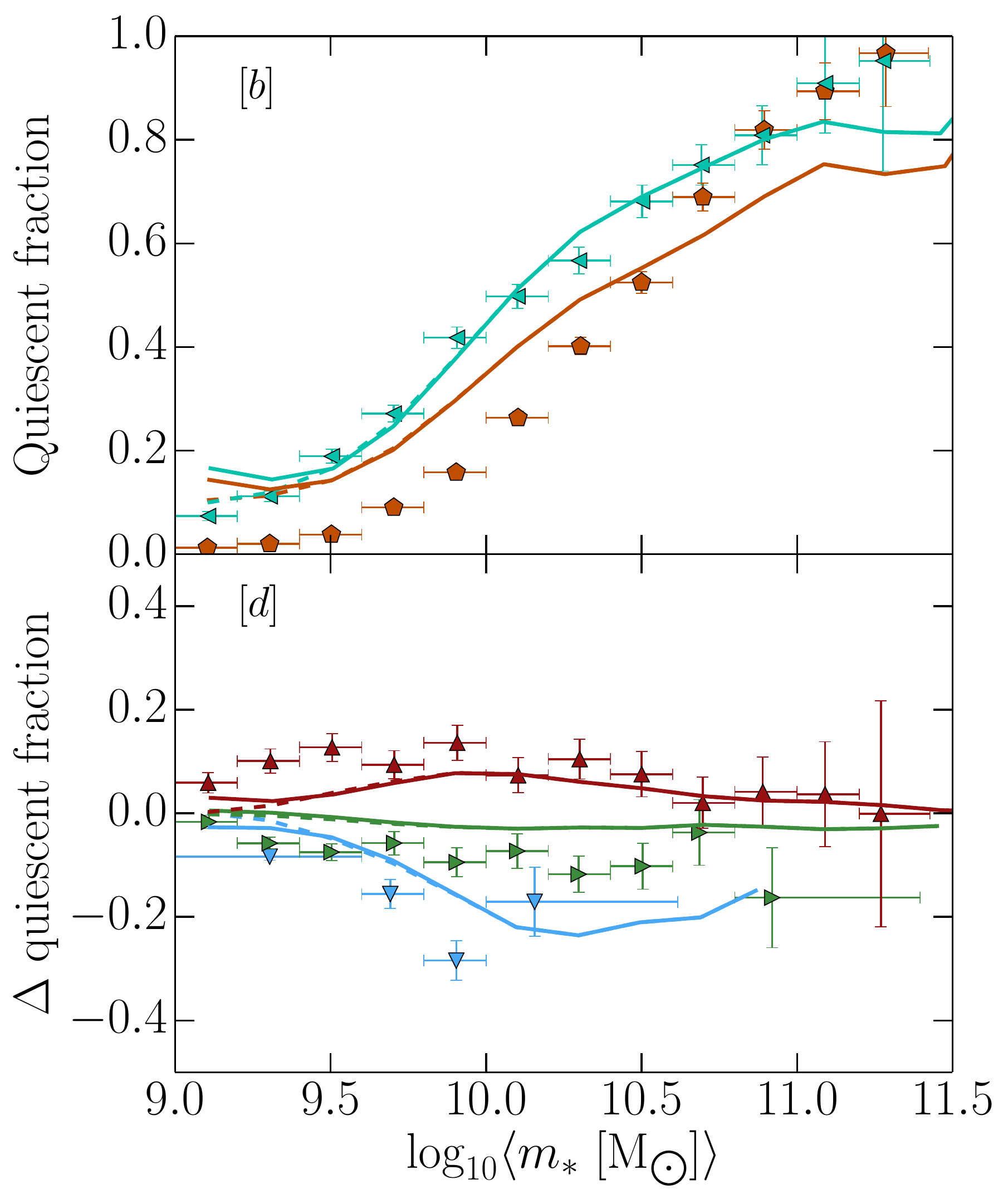}
\caption{Full-model results for {\sc Dark Sage} after cross-contaminating the central and satellite populations.  We have invoked a purity, $P$, of 0.6 and 0.8 for satellites and centrals, respectively, for all stellar and halo masses.  Observational data remain as measured.  Panels $a$, $b$, $c$, and $d$ should be compared with Figs.~\ref{fig:satcen}, \ref{fig:quiescent}, \ref{fig:satenv}, and \ref{fig:quiescent_env}, respectively.}
\label{fig:contam}
\end{figure*}

Another possible explanation for the discrepancy in the mean \HI fractions of satellites in the observations and model lies with the group-finding algorithm applied to the observational data.  \citet{campbell15} studied how well several group-finding algorithms performed in correctly identifying satellites and centrals by applying them to a mock catalogue from simulated data.  The authors found that the \citet{yang05,yang07} technique returns a population of satellites with a purity of $\sim \! 0.6$--$0.7$ for all haloes with $M_{\rm vir} \! > \! 10^{12}\,{\rm M}_{\odot}$ \citep[see fig.~6 of][]{campbell15}.  This means it is possible that 30--40 per cent of galaxies classed as satellites in our observational sample (Section \ref{sec:obs}) might actually be centrals.  This could more than account for the difference in the fraction of satellites in the observed and model samples (38.2 and 29.7 per cent, respectively -- Section \ref{ssec:satcen_full}).  While repeating the efforts of \citeauthor{campbell15} for {\sc Dark Sage} is beyond the scope of this paper, we can easily test what the model results would be after contaminating the satellite population with misidentified centrals by taking a weighted mean of the true central and satellite galaxy properties.  Using respective weights of 0.6 and 0.4 for satellites and centrals gives a mean \HI fraction as a function of stellar mass that is consistent with a satellite population with a purity of 0.6.  We have plotted this for {\sc Dark Sage} in Fig.~\ref{fig:contam}$a$, where we have also included central-galaxy results after applying a purity of 0.8.  The contaminated satellite population meets the relation for observed satellites to a remarkable degree.  While this could offer an easy out for the too-gas-poor nature of {\sc Dark Sage} satellites, cross contamination of satellites and centrals leads to a reduction in the relative gas fractions of satellites in bins of halo masses, as seen in Fig.~\ref{fig:contam}$c$.  This would, therefore, simply trade one good result for a different one.  Either way, the treatment of satellites in the semi-analytic model still requires further attention before it can reproduce all the observational results we have shown.

Our results with {\sc Dark Sage} suggest the effects of hot- and cold-gas stripping on galaxies are almost entirely separable.  Cold-gas stripping removes the highest-$j$ gas in the disc, which is at low surface density and is dominated in mass by \HInospace.  This generates a difference in the mean \HI fractions between central and satellite galaxies at fixed stellar mass (Fig.~\ref{fig:satcen}), and nicely recovers observed trends in the relative \HI fractions of satellites in different environments (probed by halo mass -- Fig.~\ref{fig:satenv}).  Molecular gas, which exists almost exclusively at the centres of these galaxies' discs, is almost unaffected by this process.  Recent CO observations of 3 satellite galaxies in the Virgo cluster by \citet{lee17} show that, while the morphology of molecular gas can be disturbed by ram pressure, there is no evidence that ram-pressure stripping affects the total molecular gas content.  For {\sc Dark Sage}, only low-$m_*$ galaxies in high-mass haloes have their molecular gas and star formation activity directly impacted by ram-pressure stripping (Fig.~\ref{fig:quiescent_env}).

On the other hand, hot-gas stripping shuts down galaxy accretion, thereby affecting the evolution of gas across the entire disc.  The replenishment of molecular gas in the galaxy is then suppressed, causing the satellites to become quiescent after consuming their available H$_2$ in star formation.  This is a major contributor behind the observed difference in quiescent fractions between satellites and centrals at fixed stellar mass (Fig.~\ref{fig:quiescent}).  Based on the relative colours, sizes, and masses of centrals and satellites, observational evidence supports a picture where hot-gas stripping is the responsible mechanism for satellite quenching \citep{bosch08}.  While showing improvement relative to earlier results from semi-analytic models, {\sc Dark Sage} still overproduces the fraction of quiescent satellites in the range $10^{10} \! \lesssim \! m_*/{\rm M}_{\odot} \! \lesssim \! 10^{11}$.  Cross-contamination of centrals and satellites in the observational data from the group finder could also be playing a role here, mind; after cross-contaminating the {\sc Dark Sage} galaxies, we find the excess of quiescent satellites at mid masses goes away (Fig.~\ref{fig:contam}$b$).  Again though, this reduces the relative strength of environment on quiescence in the model (Fig.~\ref{fig:contam}$d$).

\begin{figure}
\includegraphics[width=\linewidth]{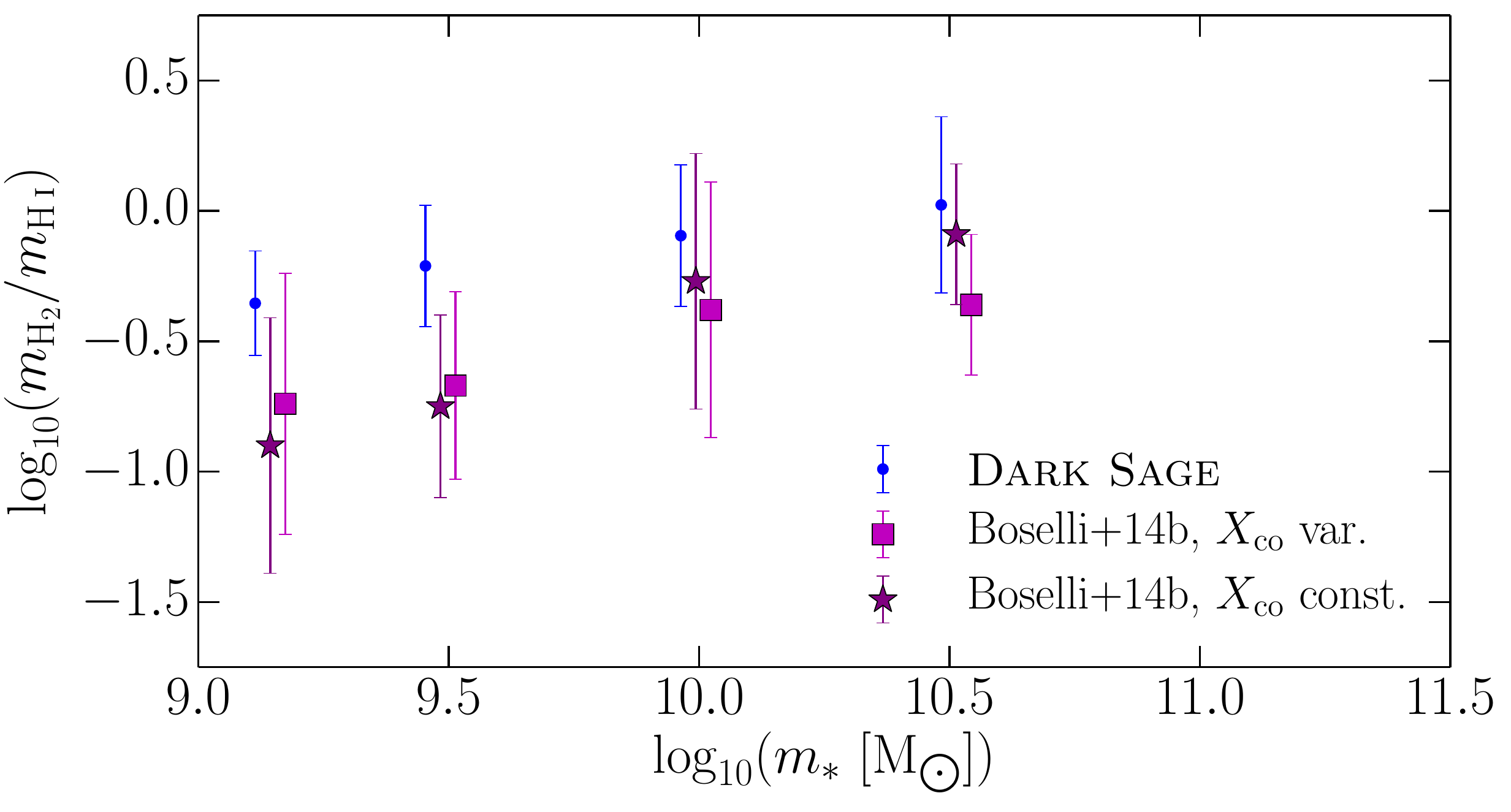}
\caption{Molecular-to-atomic hydrogen mass ratio as a function of stellar mass for galaxies at $z=0$ with $m_{\rm H_2} \! \geq \! 10^{8.6}\,{\rm M}_{\odot}$, $m_{\rm H_2} \! \geq \! 0.01\,m_{\mathrm{H}\,\LARGE\textsc{i}}$, and bulge-to-total ratios $\! < \! 0.3$ (these include both centrals and satellites).  These cuts are to approximately match the completeness limit and morphology of the compared, observed galaxies from \citet{boselli14b}.  The $x$ and $y$ positions of the points are log of each of the mean stellar mass and mean molecular-to-atomic ratio, respectively, for four bins.  Square and starred points have been manually shifted to the right by 0.03 and 0.06 dex, respectively, for the sake of clarity.    These present the same actual data, where squares employ a luminosity-dependent conversion factor between CO emission and H$_2$ mass, and starred points use a constant conversion factor.  The error bars give the standard deviation in $\log_{10}(m_{\rm H_2}/m_{\mathrm{H}\,\LARGE\textsc{i}})$ within each bin. {\sc Dark Sage} data are for the full model with the $f_{\rm H_2}(Z)$ prescription, using (sub)haloes with $N_{\rm p,max} \! \geq \! 100$.}
\label{fig:h2frac}
\end{figure}

We note that \citet{thesis} has shown that the stellar half-mass radii of disc-dominated {\sc Dark Sage} galaxies are systematically lower than the half-light radii of galaxies in the GAMA\footnote{Galaxy And Mass Assembly} survey by $\lesssim \! 0.1$\,dex \citep[cf.][]{lange16}.  Galaxies that never suffered an instability in their existence, however, meet the observed relation quite precisely.  Instabilities drive mass inwards and angular momentum outwards in {\sc Dark Sage} discs \citep{stevens16}.  Because the model lacks a detailed consideration of radial dispersion support at the centres of discs, their surface density profiles grow an exaggerated cusp.  This leads to smaller stellar sizes and higher molecular-to-atomic hydrogen mass ratios, especially at the discs' centres.  Because the integrated \HI content of the model galaxies is calibrated to meet observations \citep{brown15}, the average radius of an \HI element will be greater than in real galaxies \citep[see fig.~4 of][]{stevens16}.  This does not prevent the nominal \HI radii of the galaxies (where $\Sigma_{\mathrm{H}\,\LARGE\textsc{i}} \! = \! 1\,{\rm M}_{\odot}\,{\rm pc}^{-2}$) from agreeing with observations for fixed \HI mass, however (see Lutz et al.~in preparation). We compare the molecular-to-atomic ratio as a function of stellar mass for {\sc Dark Sage} galaxies against data presented by \citet{boselli14b} from the \emph{Herschel} Reference Survey \citep{herschel,boselli14a} in Fig.~\ref{fig:h2frac}.  Indeed, the ratios of the model galaxies are systematically higher than the observational data, regardless of whether the conversion between CO luminosity and H$_2$ mass is taken as variable or a constant.  While the \citeauthor{boselli14b} data are not volume-limited and only account for 74 late-type galaxies with detections in both \HI and CO, they are designed to be representative of average galaxies in the local Universe.  With that in mind, we encourage that while the information in Fig.~\ref{fig:h2frac} is important to consider in interpreting our results, it is unlikely to be what drives them.  Future work in improving how the centres of {\sc Dark Sage} discs are modelled, as well as larger H$_2$ surveys, will help to clarify our results.

We should be optimistic about soon converging on a detailed description of how environmental and secular processes affect the gas content and star formation activity of galaxies in a cohesive manner that explains our observations.  
Already, data from MUSE\footnote{Multi Unit Spectroscopic Explorer} \citep{muse} have been used to show the optical effects of ram pressure in exquisite detail for a small sample of galaxies \citep[e.g.][]{muse1,gasp1}, which provide clues for this picture.  
Upcoming surveys such as the ASKAP\footnote{Australian Square Kilometre Array Pathfinder} \HI All-Sky Survey \citep[also known as {\sc Wallaby};][]{wallaby} will complement these with large-number statistics of the \HI content, structure, and kinematics of environmentally influenced satellite galaxies.  
On the theory side, where detailed cosmological hydrodynamic simulations have been limited in their volume \citep[$\sim\!10^6\,{\rm Mpc}^3$ -- e.g.][]{dubois14,illustris,khandai15,schaye15}, follow-up simulations with the same well-tested physical prescriptions are now targeting Local Group analogues and massive clusters, providing more opportunity to learn about the impact of stripping processes \citep[][respectively]{sawala16,bahe17}.  
Based on the level of agreement with observations we have presented in this paper, and without any concerns regarding volume or computational efficiency, semi-analytic models are bound to continue to play an important role in this quest as well.

\section*{Acknowledgements}
We thank Claudia Lagos for comments on this paper, as well as Luca Cortese, Violeta Gonzalez-Perez, and Manodeep Sinha for helpful discussion regarding this work.  ARHS also thanks Darren Croton for continued academic support during the writing of this manuscript.

The {\sc Dark Sage} codebase is publicly available at \url{https://github.com/arhstevens/DarkSage}. This was developed from the \textsc{sage} codebase, which is also publicly available at \url{https://github.com/darrencroton/sage}.   Catalogues from both models are available at \url{https://tao.asvo.org.au/tao/}.

\bibliographystyle{mnras}

\end{document}